\documentclass[twocolumn]{aastex631}

\usepackage{amsmath}
\usepackage{longtable}
\usepackage{float}

\shorttitle{Adiabatic Mass Loss in Binary Stars. V.}
\shortauthors{Ge et al.}
\graphicspath{{./}}
\begin{document}

\title{Adiabatic Mass Loss in Binary Stars. V. Effects of Metallicity and Nonconservative Mass Transfer -- Application in High Mass X-ray Binaries}

\correspondingauthor{Hongwei Ge}
\email{gehw@ynao.ac.cn}

\author[0000-0002-6398-0195]{Hongwei Ge}
\affiliation{Yunnan Observatories, Chinese Academy of Sciences,
	Kunming, 650216, People's Republic of China}
\affiliation{Key Laboratory for Structure and Evolution of Celestial Objects, 
	Chinese Academy of Sciences, Kunming 650216, People's Republic of China}
\affiliation{International Centre of Supernovae, Yunnan Key Laboratory,
	Kunming 650216, People's Republic of China}
\affiliation{University of Chinese Academy of Sciences, Beijing 100049, People's Republic of China}
%\email{gehw@ynao.ac.cn}

\author[0000-0002-1556-9449]{Christopher A Tout}
\affiliation{Institute of Astronomy, The Observatories, University of Cambridge, Madingley Road, Cambridge, CB3 0HA, UK}
%\email{cat@ast.cam.ac.uk}

\author[0000-0001-5284-8001]{Xuefei Chen}
\affiliation{Yunnan Observatories, Chinese Academy of Sciences,
	Kunming, 650216, People's Republic of China}
\affiliation{Key Laboratory for Structure and Evolution of Celestial Objects, 
	Chinese Academy of Sciences, Kunming 650216, People's Republic of China}
\affiliation{International Centre of Supernovae, Yunnan Key Laboratory,
		Kunming 650216, People's Republic of China}
\affiliation{University of Chinese Academy of Sciences, Beijing 100049, People's Republic of China}
%\email{cxf@ynao.ac.cn}

\author[0000-0003-3116-5038]{Song Wang}
\affiliation{Key Laboratory of Optical Astronomy, National Astronomical Observatories, Chinese Academy of Sciences, Beijing 100101, People's Republic of China}
\affiliation{Institute for Frontiers in Astronomy and Astrophysics, Beijing Normal University, Beijing 102206, People's Republic of China}

\author[0000-0003-4829-6245]{Jianping Xiong}
\affiliation{Yunnan Observatories, Chinese Academy of Sciences,
	Kunming, 650216, People's Republic of China}
\affiliation{Key Laboratory for Structure and Evolution of Celestial Objects, 
	Chinese Academy of Sciences, Kunming 650216, People's Republic of China}
\affiliation{International Centre of Supernovae, Yunnan Key Laboratory,
	Kunming 650216, People's Republic of China}

\author[0009-0001-3638-3133]{Lifu Zhang}
\affiliation{Yunnan Observatories, Chinese Academy of Sciences, Kunming, 650216, People's Republic of China}
\affiliation{University of Chinese Academy of Sciences, Beijing 100049, People's Republic of China}

\author{Qingzhong Liu}
\affiliation{Key Laboratory of Dark Matter and Space Astronomy, Purple Mountain Observatory, ChineseAcademy of Sciences, Nanjing,  People's Republic of China}

%\author{Ronald F Webbink}
%\affiliation{University of Illinois at Urbana-Champaign, 1002 W Green St, Urbana, 61801, USA}
%\email{rwebbink@illinois.edu}

\author[0000-0001-9204-7778]{Zhanwen Han}
\affiliation{Yunnan Observatories, Chinese Academy of Sciences,
	Kunming, 650216, People's Republic of China}
\affiliation{Key Laboratory for Structure and Evolution of Celestial Objects, 
	Chinese Academy of Sciences, Kunming 650216, People's Republic of China}
\affiliation{International Centre of Supernovae, Yunnan Key Laboratory,
		Kunming 650216, People's Republic of China}
\affiliation{University of Chinese Academy of Sciences, Beijing 100049, People's Republic of China}
%\email{zhanwenhan@ynao.ac.cn}

\begin{abstract}
Binary stars are responsible for many unusual astrophysical phenomena, including some important explosive cosmic events. The stability criteria for rapid mass transfer and common-envelope evolution are fundamental to binary star evolution. They determine the mass, mass ratio, and orbital distribution of systems such as X-ray binaries and merging gravitational-wave sources. We use our adiabatic mass-loss model to systematically survey metal-poor and solar-metallicity donor thresholds for dynamical timescale mass transfer. The critical mass ratios $q_\mathrm{ad}$ are systematically explored and the impact of metallicity and nonconservative mass transfer are studied. For metal-poor radiative-envelope donors, $q_\mathrm{ad}$ are smaller than those for solar-metallicity stars at the same evolutionary stage. However, $q_\mathrm{ad}$ do the opposite for convective-envelope donors. Nonconservative mass transfer significantly decreases $q_\mathrm{ad}$ for massive donors. This is because it matters how conservative mass transfer is during the thermal timescale phase immediately preceding a delayed dynamical mass transfer. We apply our theoretical predictions to observed high-mass X-ray binaries that have overfilled their Roche lobes and find a good agreement with their mass ratios. Our results can be applied to study individual binary objects or large samples of binary objects with binary population synthesis codes.
\end{abstract}


%\turnoffedit
%\turnoffediting
\section{Introduction} 
\label{sec:intro}

Binary stars lie at the heart of many vital astrophysical phenomena \citep[e.g.,][]{2023pbse.book.....T}. Different types of explosive cosmic events are produced by binary star objects, such as double black holes \citep[DBHs, e.g.,][]{2022LRR....25....1M}, double neutron stars \citep[DNSs, e.g.,][]{2023ApJ...955..133G}, double white dwarfs \citep[DWDs, e.g.,][]{2024ResPh..5907568L}, type Ia supernovae \citep[e.g.,][]{2023RAA....23h2001L} and X-ray binaries \citep[e.g.,][]{2023pbse.book.....T}. The study of binary evolution brings new opportunities for insights into fundamental physics \citep[e.g.,][]{2023LRR....26....2A} such as testing general relativity, production of heavy chemical elements, and discovering new double compact systems in the next few decades. Despite the importance of binary evolution, there remain unsolved fundamental questions of particular importance are mass transfer stability \citep{1984ApJ...277..355W} and common envelope (CE) evolution \citep{1976IAUS...73...75P}. 

A series of recent study indicate that mass transfer in massive binary systems \citep{2015ApJ...812...40G} and binaries with red/asymptotic giant branch (R/AGB) compaions \citep{2020ApJ...899..132G,2020ApJS..249....9G} is more stable than previously found. With these mass transfer thresholds as physical inputs, an isolated binary BH formation scenario consisting of a stable mass transfer during the second mass transfer phase (instead of CE) has been gaining popularity in the gravitational wave (GW) community \citep{2019MNRAS.490.3740N}. Besides the mass transfer physics, the supernova natal kicks \citep{2021MNRAS.500.1380M} are also critical for massive binary evolution. Hence, mass transfer is probably active in some systems in eccentric orbits. As for WD binaries, mostly formed from R/AGB progenitors, a first stable mass transfer plus a second CE channel may dominate the formation of DWDs \citep{2023A&A...669A..82L}. These objects are low-frequency GW sources detectable with Tianqin and LISA detectors \citep{2016CQGra..33c5010L,2017arXiv170200786A}. 

Many observed stellar phenomena, including binaries, are dominated by metal-poor environments \citep{2023ApJ...945....7G}. Examples include blue stragglers \citep{1953AJ.....58...61S}, blue metal-poor stars \citep{2000AJ....120.1014P}, Galactic halo stars \citep{2018ApJS..238...16L}, metal-poor thick disk stars \citep{2021MNRAS.501.4917W}, ultra-luminous X-ray sources (ULXs,  \citealt{2009MNRAS.400..677Z}), partially stripped-envelope stars \citep{2022A&A...662A..56K}, massive stellar-mass black holes (BHs) and their mergers \citep{2021MNRAS.504..146V}, and metal-poor inner Galactic stars \citep{2024MNRAS.530.3391A}. Nonconservative mass transfer in binary evolution, mass and angular momentum loss, plays an important role in binary evolution \citep{2024PrPNP.13404083C}. It may substantially impact predictions about the production rates of various transients, such as X-ray binaries and the progenitors of coalescing compact binaries \citep{2023ApJ...958..138W}. 

We extend our systematic studies of stars to metal-poor donors and nonconservative mass transfer. With our adiabatic mass loss model, we only provide critical mass ratio thresholds \citep{2010ApJ...717..724G} assuming conservative mass transfer in the first place. However, the advantage is that the donor response to rapid mass transfer is independent of the binary orbital evolution. So our studies can easily resolve the nonconservative mass transfer cases. Our preliminary results have been included in binary population synthesis and applied to study objects such as DWDs \citep{2023A&A...669A..82L}. After applied the new thresholds, the results support the observational DWD merger rate distribution per Galaxy and the space density of DWDs in the Galaxy.

In Section \ref{sec-method} we describe methods to build an adiabatic mass-loss model and solve for the critical mass ratio under conservative and nonconservative mass transfer. Using massive $20\,M_\odot$ stars as examples, in Section \ref{sec-20m} we study the impacts of metallicity and nonconservative mass transfer on the critical mass ratios for dynamical timescale mass transfer in metal-poor $Z=0.001$ and solar-metallicity $Z=0.02$ stars \footnote{Recent studies indicate a lower metallicity in the solar atmosphere \citep[e.g.,][]{2009ARA&A..47..481A}}. In Section \ref{sec-result} we present our model grids and thresholds for unstable mass transfer. In Section \ref{sec-hmxb} we collect observed high-mass X-ray binaries (HMXBs) with known binary parameters and compare mass ratios and orbital periods with our theoretical predictions. In Sections \ref{sec-discussion} and \ref{sec-summary} we discuss and summarize our systematic studies of the thresholds for dynamical timescale mass transfer.

\section{Methods}
\label{sec-method}

We use our adiabatic mass-loss model to study the responses of donor stars undergoing rapid mass transfer. \citet{2010ApJ...717..724G} decribed the detailed method and numerical implementation. The stellar evolution code is updated and extended by \citet{2015ApJ...812...40G,2020ApJ...899..132G,2023ApJ...945....7G} and is based on the Cambridge Stars Code \citep{1995MNRAS.274..964P} originally developed by \citet{1971MNRAS.151..351E}.

We extend our previous studies of dynamical timescale mass transfer thresholds for nonconservative cases. This extention is essential for the {\it delayed dynamical instability} of donor stars with a deep radiative envelope. We first recall the key points to calculate the critical initial mass ratio $q_\mathrm{ad}$ for the dynamical timescale mass transfer. The classical method \citep{1985ibs..book...39W} is to compare the slope of the mass-radius relation of both the stellar radius $R$ and the Roche radius $R_\mathrm{L}$, such as equations 2 to 11 of \citet{2010ApJ...717..724G}. The slopes are expressed by the so-called mass-radius exponents,
\begin{equation}
	\zeta_\mathrm{ad} \equiv \left( \frac{d \log R}{d \log M}\right)_\mathrm{ad},
\end{equation}
and 
\begin{equation}
	\zeta_\mathrm{L} \equiv  \frac{d \log R_\mathrm{L}}{d \log M},
\end{equation}
where $\zeta_\mathrm{ad}$ $<$ $\zeta_\mathrm{L}$ implies instability on a dynamical timescale mass transfer and $\zeta_\mathrm{ad} \geq \zeta_\mathrm{L}$ implies stable mass transfer. However, runaway mass transfer increases gradually and some donors may have a loose and low-density envelope. So, instead of using the surface radius $R$ of the donor star, we use an inner radius $R_\mathrm{KH}$ to calculate the mass-radius exponent, $\zeta^\mathrm{KH}_\mathrm{ad} =d \ln R_\mathrm{KH}/ d \ln M |_\mathrm{ad}$. This innner radius $R_\mathrm{KH}$ determines when the mass-loss rate $\dot{M}$ (see A9 of \citealt{2010ApJ...717..724G}) reaches a thermal timescale rate $\dot{M}_\mathrm{KH} = M/t_\mathrm{KH}$.

We assume a donor star overfills its Roche lobe at some initial mass and radius (the $n$th model from evolutionary sequences counting from the zero-age main-sequence, ZAMS). The critical initial mass ratio $q_\mathrm{i} \equiv M_\mathrm{donor}/M_\mathrm{accretor} = M_\mathrm{d}/M_\mathrm{a}$ for dynamical timescale mass transfer is to be determined. We define a mass function $\mu = M_\mathrm{d}/(M_\mathrm{d} + M_\mathrm{a}) = q/(1 + q)$ which lies between 0 and 1. For the adiabatic mass-loss sequences, we enumerate models by $b \in{\{1,2,...,B\}}$. Initially, $b=1$, the initial mass is $M_\mathrm{i}$ and radius $R_\mathrm{i}$. The initial separation $A_\mathrm{i}$ is such that $R_\mathrm{i} = R_\mathrm{L}$, the Roche-lobe radius of the donor. For model $b$, the remaining total mass $M_\mathrm{b}$ and the innner radius $R^{b}_\mathrm{KH}$ are found by our adiabatic mass-loss \citep{2010ApJ...717..724G}. The mass-radius exponent $\zeta^\mathrm{KH}_\mathrm{ad}$ can thus be found with models $b-1$ and $b$. It is independent of whether the mass transfer is conservative or not. To find the critical initial mass ratio, we need to know the mass-radius exponent $\zeta_\mathrm{L}$, which depends on how conservative the mass transfer is.

The inner Lagrangian Roche-lobe radius $R_\mathrm{L}$ of the donor \citep{1983ApJ...268..368E} can be expressed by 
\begin{equation}
 \frac{R_\mathrm{L}}{A} = r_\mathrm{L}(q) = \frac{0.49 q^{2/3}}{0.6q^{2/3}+ \ln (1+q^{1/3})}.
 \label{RL1}
\end{equation}
Differentiating equation (\ref{RL1}) we have
\begin{equation}
	\frac{d \ln r_{\mathrm{L}}}{d \ln q}=\frac{1}{3}\left[\frac{2 \ln \left(1+q^{1 / 3}\right)-q^{1 / 3} /\left(1+q^{1 / 3}\right)}{0.6 q^{2 / 3}+\ln \left(1+q^{1 / 3}\right)}\right].
\end{equation}
According to \citet{2020ApJS..249....9G}, the outer Lagrangian Roche-lobe radius of the donor $R^\mathrm{out}_\mathrm{L} = r^{\mathrm{out}}_{\mathrm{L}}(q) A $ can be found from 
\begin{equation}
\begin{aligned}
	r^{\mathrm{out}}_{\mathrm{L}}(q)= & \,r_{\mathrm{L}}(q)+\left[0.179+0.01\left(\frac{q}{1+q}\right)\right] \\
	& \times\left(\frac{q}{1+q}\right)^{0.625} \text { for } q \leqslant 1,
\end{aligned}
\end{equation}
and 
\begin{equation}
\begin{aligned}
	r^{\mathrm{out}}_{\mathrm{L}}(q) & =r_{\mathrm{L}}(q)+\left[0.179+0.01\left(\frac{q}{1+q}\right)\right. \\
	& \left.-\,0.025\left(\frac{q-1}{q}\right)\right]\left(\frac{q}{1+q}\right)^{0.625} q^{-0.74} \text { for } q \geqslant 1 .
\end{aligned}
\end{equation} 
In terms of Lograngian surfaces, the physical radius of the donor outer lobe is $R_{\rm L_3}$ for $q>1$ and $R_{\rm L_2}$ for $q<1$.

For conserved mass and angular momentum, \citet{2023ApJ...945....7G} found
\begin{equation}
\frac{R_{\mathrm{L}}\left(\mu_{b}\right)}{R_{\mathrm{i}}}=\frac{r_{\mathrm{L}}\left(\mu_{b}\right)}{r_{\mathrm{L}}\left(\mu_{\mathrm{i}}\right)}\left(\frac{\mu_{\mathrm{i}}}{\mu_{b}}\right)^{2}\left(\frac{1-\mu_{\mathrm{i}}}{1-\mu_{b}}\right)^{2},
\end{equation} 
where $\mu_{b}=\mu_{\mathrm{i}} M_{b} / M_\mathrm{i}$ and $\mu=q /\left( 1+q \right)$. For nonconservative mass transfer, a fraction $\alpha$ of the total mass lost by the donor is transferred to its companion. So a fraction $1-\alpha$ of the donor mass lost carries away the donor's specific angular momentum. We further suppose that a faction $\beta$ of the transferred mass is accreted so that a faction ($1-\beta$) of the transferred mass is ejected from the accretor with $\gamma$ times its specific angular momentum. So for mass-loss model $b$, the mass ratio
\begin{equation}
 q = \frac{M_b}{M_{\rm i}/q_{\rm i}+\alpha \beta (M_{\rm i}-M_b) },
\end{equation}
and the separation varies as
\begin{equation}
	\begin{aligned}
	\frac{A_b}{A_{b-1}} = &   1+ \frac{2 (M_{b-1}-M_b)}{M_b} \times \\
	&\left[\alpha (1-q \beta)-  \frac{q}{1+q}(\gamma q+0.5) (1-\alpha \beta) \right] .
	\end{aligned}
\end{equation}
When mass is only lost from the accretor and carries away its specific angular momentum, we have $\alpha =1$ and $\gamma = 1$. We set $\beta =1$ (fully conservative), 0.5 (nonconservative), and 0 (fully nonconservative) in this study. Starting from an initial guess, $\mu_\mathrm{i} = 0.5$, we use a bisection method to calculate the initial mass function, $\mu_b$, satisfying both $R^b_\mathrm{L} = R^b_\mathrm{KH}$ and $\zeta^b_\mathrm{L}= \zeta^b_\mathrm{KH}$. By tracing $b = 1$ to $b = B$ for the whole adiabatic mass loss process, we can get the minimum value of $\mu_\mathrm{min}$. So, the critical initial mass ratio is calculated finally with $q_\mathrm{ad} = \mu_\mathrm{min}/(1-\mu_\mathrm{min})$.

%For the outer critical surface when $q < 1$,
%\begin{equation}
%	\frac{d \ln r^\mathrm{out}_\mathrm{L}}{d \ln q}=\frac{q}{r^\mathrm{out}_\mathrm{L}} \frac{d r^\mathrm{out}_\mathrm{L}}{d q},
%\end{equation}
%with
%\begin{equation}
%	\begin{aligned}
%		\frac{d r^\mathrm{out}_\mathrm{L}}{d q}= &\frac{r_{\mathrm{L}}}{q} \frac{d \ln r_{\mathrm{L}}}{d \ln q}+\frac{1}{(1+q)^{2}}\left(\frac{q}{1+q}\right)^{0.625}\\
%		&\times \left[0.128125+\frac{0.111875}{q}\right] .
%	\end{aligned}
%\end{equation}
%And when $q > 1$,
%\begin{equation}
%	\frac{d \ln r^\mathrm{out}_\mathrm{L}}{d \ln q}=\frac{q}{r^\mathrm{out}_\mathrm{L}} \frac{d r^\mathrm{out}_\mathrm{L}}{d q},
%\end{equation}
%where
%\begin{equation}
%	\begin{aligned}
%		\frac{d r^\mathrm{out}_\mathrm{L}}{d q}= & \frac{r_{\mathrm{L}}}{q} \frac{d \ln r_{\mathrm{L}}}{d \ln q}+\left(\frac{q}{1+q}\right)^{0.625} q^{-0.74}\left\{\frac{0.01}{(1+q)^{2}}- \right. \\ 
%		&\frac{0.025}{q^{2}} +\left[\frac{0.625}{q(1+q)}-\frac{0.74}{q}\right]\times \\
%		&\left. \left[0.179+0.01\left(\frac{q}{1+q}\right)-0.025\left(\frac{q-1}{q}\right)\right]\right\}.
%	\end{aligned}
%\end{equation}

Similar to \citet{2015ApJ...812...40G,2020ApJ...899..132G} we  build parallel donor star sequences with isentropic envelopes alongside standard donor stars with standard convective envelopes. In such stars, convective envelopes are replaced by fully isentropic envelopes. With this the limitations of an adiabatic approximation in a thin layer just below the photosphere are overcome. For such stars we calculate critical initial mass ratios $\tilde{q}_\mathrm{ad}$.

\begin{figure}[htb!]
	\centering
	\includegraphics[scale=0.25]{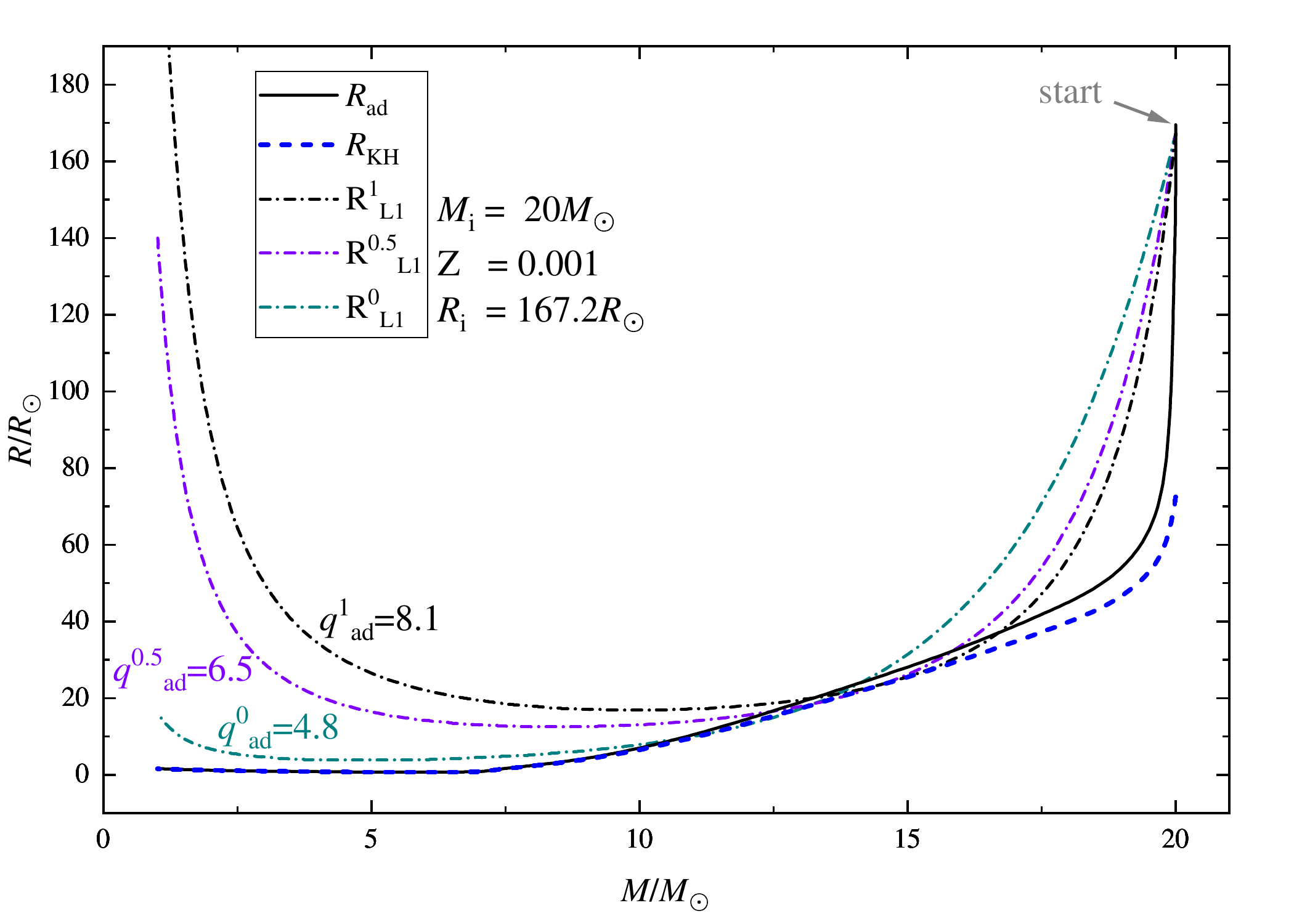}
	\caption{Example critical mass ratios $q_\mathrm{ad}$ for conservative and nonconservative mass transfer. The initial parameters are labeled. The black solid line marks the radius response during the adiabatic mass loss process. The blue dashed line is the inner radius $R_\mathrm{KH}$. The dash-dotted lines are the Roche lobe radii under conservative (black), semiconservative (violet), and fully nonconservative (cyan) mass transfer. }
	\label{fig-20m-qmethod}
\end{figure}

We demonstrate our method for a metal-poor star of $20\,M_\odot$ and  $167.2\,R_\odot$ in Figure\,\ref{fig-20m-qmethod}. The Roche lobe radius becomes much smaller when the mass transfer is more nonconservative. This allows a more shallow decent of Roche lobe radius for nonservative mass transfer. Hence, critical mass ratios $q_\mathrm{ad}$ become smaller rather than larger for nonservative mass transfer.

\section{Massive Stars of $20\,M_\odot$ with Varying Metallicities}
\label{sec-20m}

\begin{figure}[b!]
	\centering
	\includegraphics[scale=0.25]{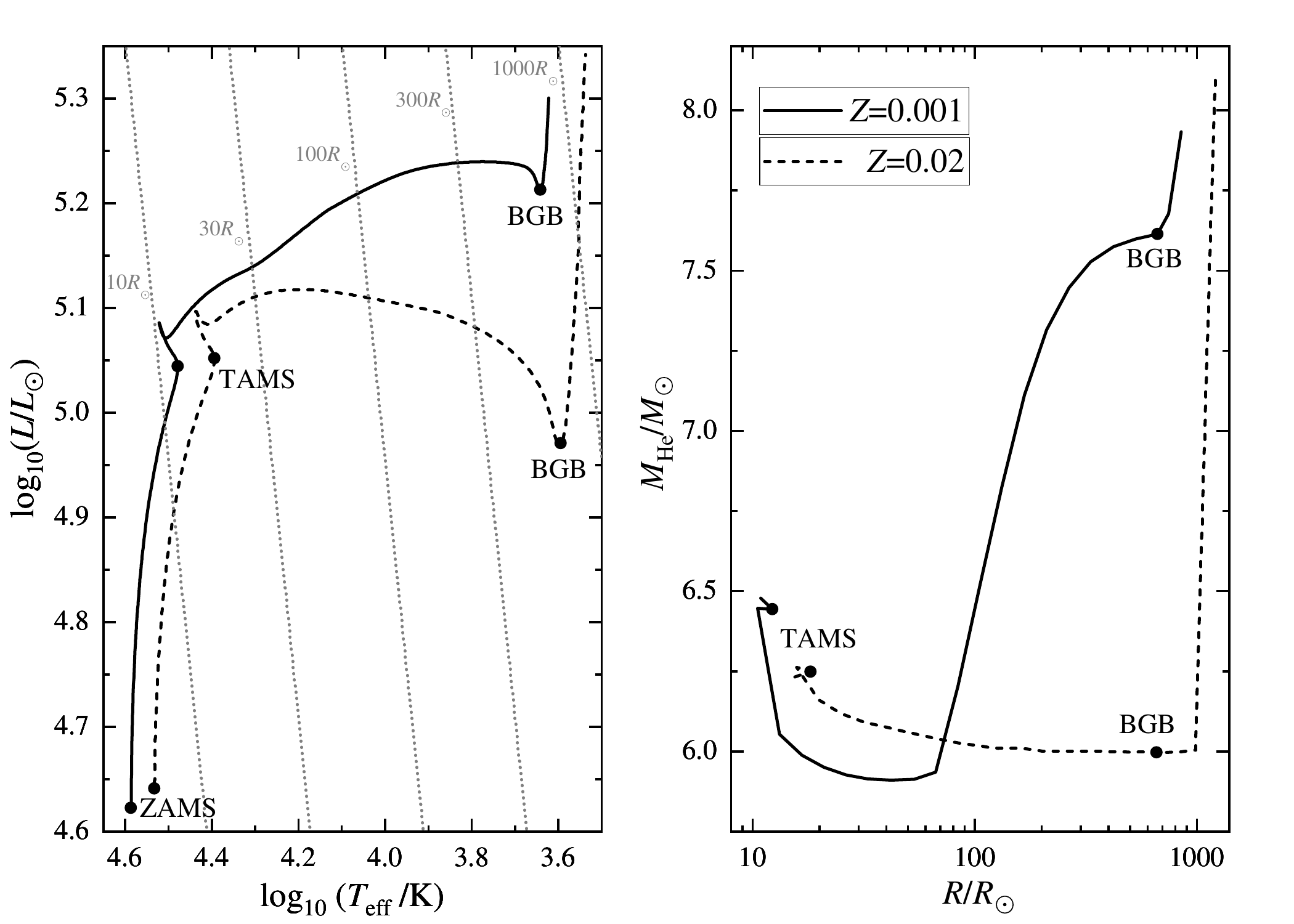}
	\caption{The left panel shows the evolution for $20\,M_\odot$ stars with different metallicities in the Hertzsprung-Russell diagram. The right panel presents their helium core masses $M_\mathrm{He}$, where core hydrogen mass fraction first reaches 0.15, as a function of their radii. Solid and dashed lines are stars with metallicities of $Z=0.001$ and $Z=0.02$.}
	\label{fig-20m-hrd}
\end{figure}

We choose $20\,M_\odot$ massive donor stars as examples to demonstrate the impacts of both metallicity and mass transfer physics. A metal-poor star behaves as if more massive than a similar mass solar-metallicity star(Figure\,\ref{fig-20m-hrd}). Such behavior is similar for intermediate-mass donor stars \citep{2023ApJ...945....7G}. A metal-poor massive star has a slightly higher central temperature $T_\mathrm{c}$ and develops a larger helium core mass $M_\mathrm{He}$ at the end of the main sequence. However, the convective core moves off-center and shrinks for a while until helium ignites in the center ofer which $M_\mathrm{He}$ increases to exceed that of a $Z=0.02$ star of the same radius (Figure\,\ref{fig-20m-hrd}).

\begin{figure}[htb!]
	\centering
	\includegraphics[scale=0.25]{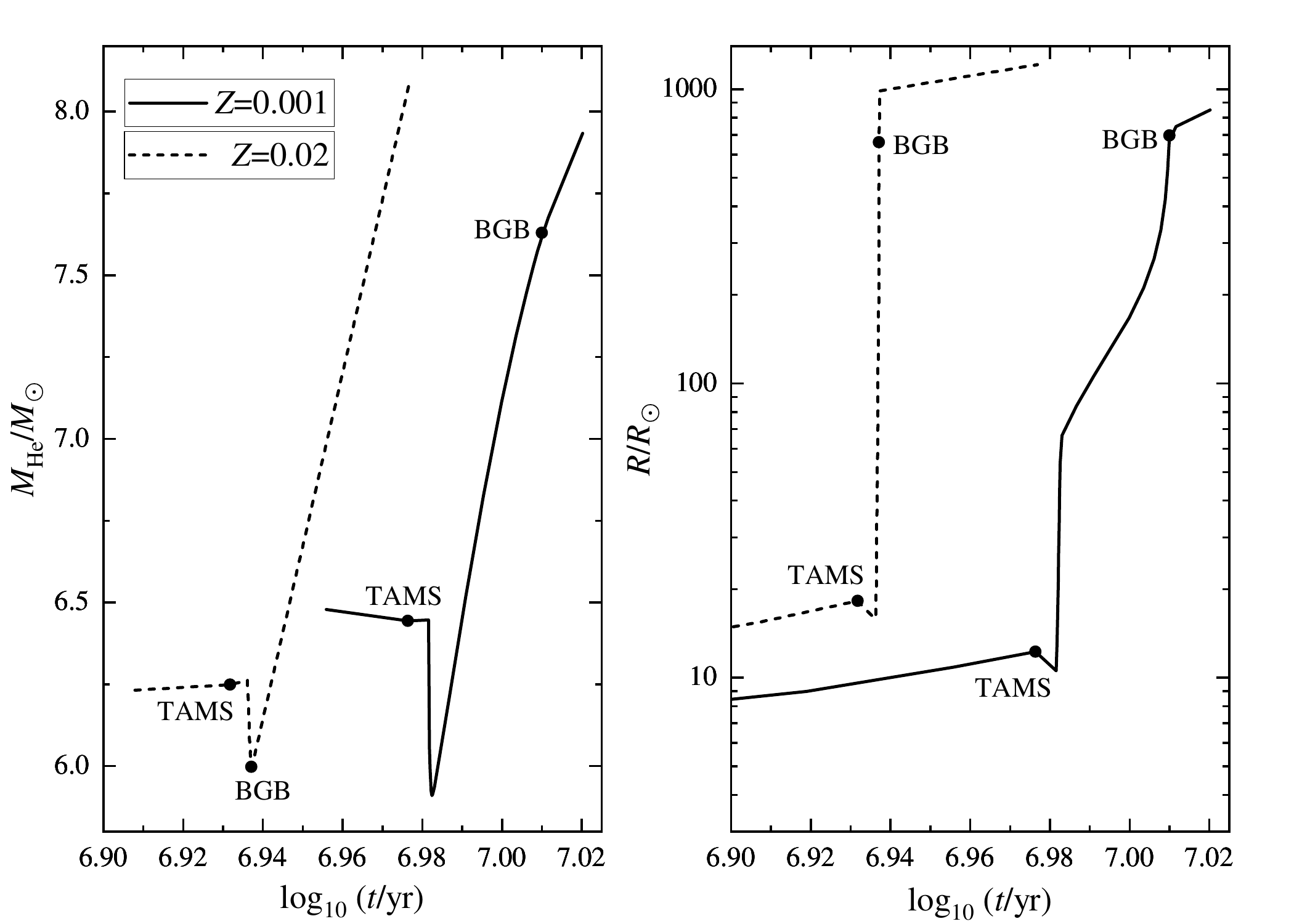}
	\caption{The impact of metallicity on helium core mass $M_\mathrm{He}$ (left panel) and the radius $R$ (right panel) as functions of a $20\,M_\odot$ star's age. Solid and dashed lines are $Z=0.001$ and $Z=0.02$ stars.}
	\label{fig-20m-mc-r}
\end{figure}

The stellar radius $R$ can be used to represent the evolutionary state of a star and because it determines when Roche-lobe overflow begins. The initial orbital period $P_\mathrm{orb}$, or separation $A$, of a binary system only weakly depends on mass ratio but strongly on donor radius \citep[][Section 2]{2015ApJ...812...40G}. Metal-poor ($Z=0.001$) massive stars have longer core hydrogen burning lifetime \citep[e.g.,][]{1992A&AS...96..269S} and smaller radii $R$ than those of same mass but $Z=0.02$ star (Figure\,\ref{fig-20m-mc-r}). 

\begin{figure}[htb!]
	\centering
	\includegraphics[scale=0.25]{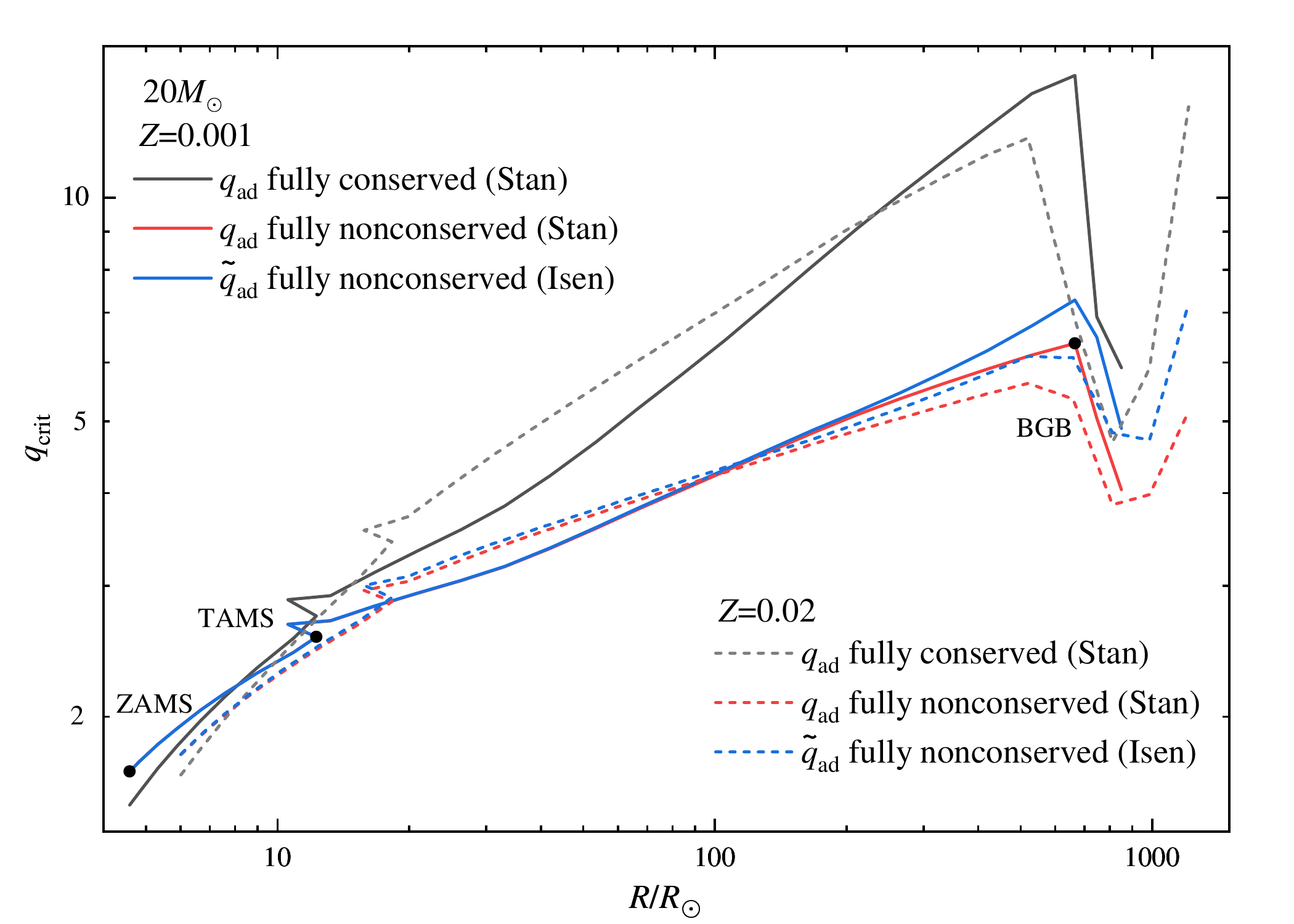}
	\caption{Critical initial mass ratios $q_\mathrm{ad}$ as functions of stellar radius for $20\,M_\odot$ models. The solid and dashed lines correspond to $Z=0.001$ and $Z=0.02$ stellar models. Red and blue lines correspond to models with standard mixing-length envelopes $q_\mathrm{ad}$ and artificially isentropic convective envelopes $\tilde{q}_\mathrm{ad}$ for fully nonconservative mass transfer. Gray lines show stellar models with standard envelopes $q_\mathrm{ad}$ for fully conservative mass transfer. }
	\label{fig-20m-qad}
\end{figure}

\citet{2020ApJ...899..132G} presented critical mass ratios $q_\mathrm{ad}$  for $0.1 \leq M/M_\odot \leq 100$ stars with solar metallicity. They confirmed that donor stars with radiative and convective envelopes behave differently during adiabatic mass loss. Intermediate-mass \citep{2020ApJ...899..132G,2023ApJ...945....7G} and massive stars  \citet{2020ApJ...899..132G} while on the main sequence (MS), Hertzsprung gap (HG), and just before the base of the giant branch (BGB) have a similar structure, including an outer radiative envelope. Although the massive star is complicated by multiple convection zones, its radiative envelope dominates behavior under adiabatic mass loss. Such donor stars are predicted to undergo delayed dynamical instability \citep{1987ApJ...318..794H,2010ApJ...717..724G} if the initial mass ratio $q_\mathrm{i}$ is larger than $q_\mathrm{ad}$. We find that the critical mass ratios $q_\mathrm{ad}$ and $\tilde{q}_\mathrm{ad}$ increase with the radius, as the star evolves from the ZAMS to the BGB (Figure\,\ref{fig-20m-qad}). The convection envelope dominates once a massive star becomes a red GB or a red supergiant (RSG). However, the thermal timescale becomes very short for convective envelope giants. So the threshold for thermal timescale mass loss through the outer-Lagrangian point $q^\mathrm{Lout}_\mathrm{th}$ is likely to become more strict compared to $q_\mathrm{ad}$ \citep{2020ApJS..249....9G}. In other words, binary systems, containing such kind of donors, with initial mass ratios $q^\mathrm{Lout}_\mathrm{th} < q_\mathrm{i} < q_\mathrm{ad}$ may enter a common envelope like phase.

After the mass of a star, its metallicity is the second most important parameter for its structure and evolution. However, metallicity seems only to shift the critical mass ratios $q_\mathrm{ad}$ of massive stars (same mass) in radius (Figure\,\ref{fig-20m-qad}). The critical mass ratio trend looks quite similar for $20\,M_\odot$ massive stars of different metallicities. As we described in the previous section, the donor response to adiabatic mass loss is independent of the orbital evolution of a rapidly mass-transferring binary system. However, how much mass and angular momentum is lost from the binary system significantly affects the Roche lobes of two components and the orbital period. So, before entering a delayed dynamical instability, conservative and nonconservative mass transfer alter the initial critical mass ratio $q_\mathrm{ad}$ dramatically (Figure\,\ref{fig-20m-qad}). The critical mass ratios of evlved donors of different metallicities decrease significantly for nonconservative mass transfer. These reduced critical initial mass ratios are caused by a longer thermal timescale mass transfer before entering the delayed unstable process. Surprisingly, the differences between fully nonconservative $q_\mathrm{ad}$ (red) and $\tilde{q}_\mathrm{ad}$ (blue) along to metallicity are not significant (Figure\,\ref{fig-20m-qad}).

\section{Model Grids and Thresholds for Unstable Mass Transfer}
\label{sec-result}

Previously \citet{2020ApJ...899..132G} studied the adiabatic responses of Population I (Z = 0.02) stars spanning a full range of stellar mass ($0.10$ to $100\,M_\odot$) and all evolutionary stages. \citet{2023ApJ...945....7G} extended the study to metal-poor donor stars. Here we also include nonconservative mass loss. This is essential for binary population synthesis studies of compact binaries with WD, NS, or BH companions.

\begin{figure}[t!]
	\centering
	\includegraphics[scale=0.25]{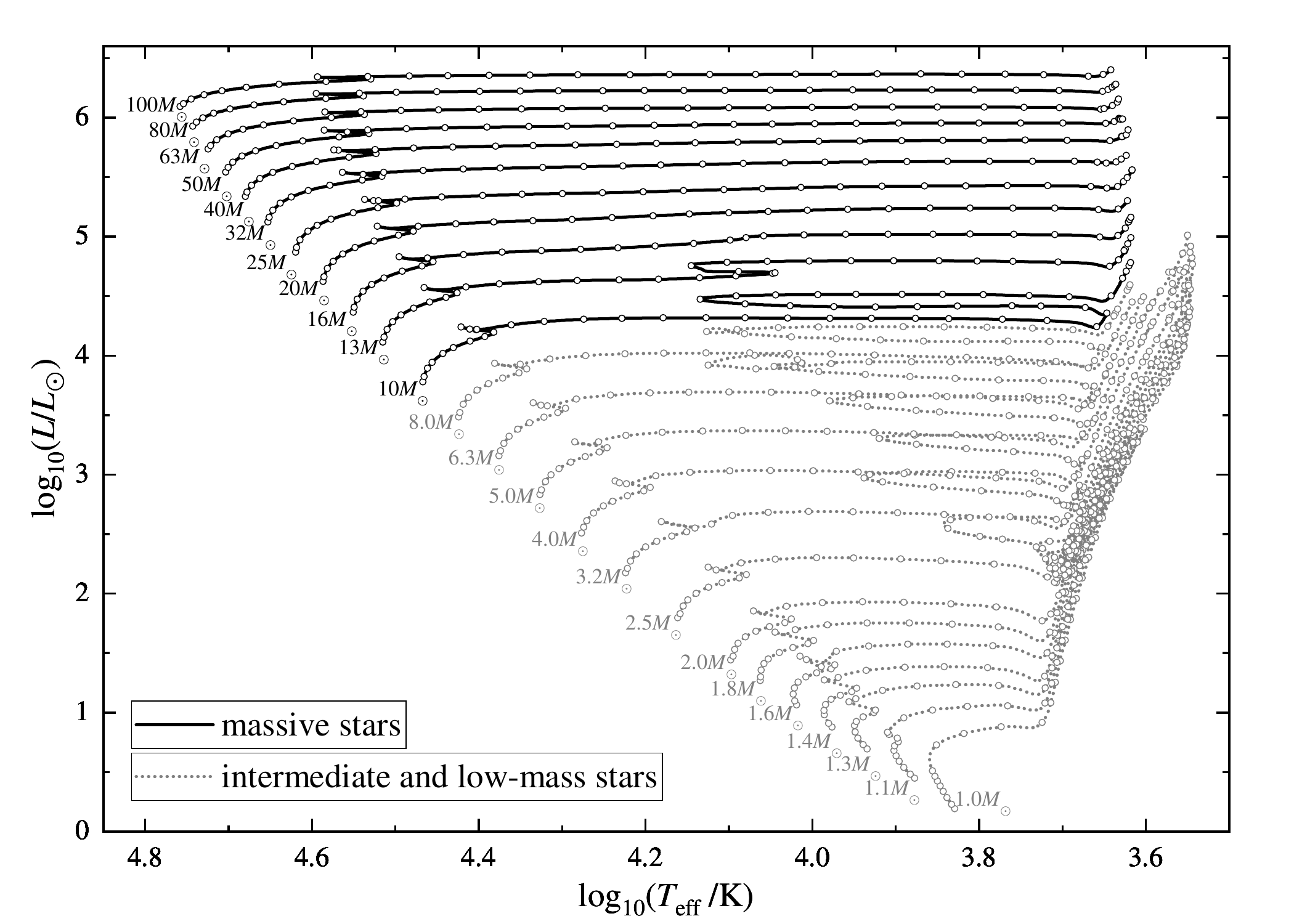}
	\caption{The evolution of metal-poor ($Z=0.001$) stars. Black solid lines indicate massive donor stars, while gray solid lines are low- and intermediate-mass stars. Some of the intermediate and low-mass stars were studied by \citet{2023ApJ...945....7G}. We also extend the donor spaces in this study. Open circles are placed at $\Delta \log_{\rm 10} R = 0.1$ from the ZAMS.}
	\label{fig-hrd}
\end{figure}

We cover a mass range from $1$ to $100\,M_\odot$ (Figure\,\ref{fig-hrd}). We present the initial physical properties of different donor stars and their thresholds for dynamical timescale mass transfer in Appendix\,\ref{sec:alltab} and online. Tables\,\ref{tab-interior} and \ref{tab-global} list interior and global physical properties of metal-poor donor stars. We only provide a $20\,M_\odot$ stellar sequence in this print version but list all the data in a machine-readable version. Table\,\ref{tab-threshold} lists thresholds for conservative and nonconservative dynamical timescale mass transfer for metal-poor stars. Table\,\ref{tab-threshold2} extends the thresholds for nonconservative dynamical timescale mass transfer in solar-metallicity stars (from $0.1$ to $100\,M_\odot$).

\begin{figure}[htb!]
	\centering
	\includegraphics[scale=0.25]{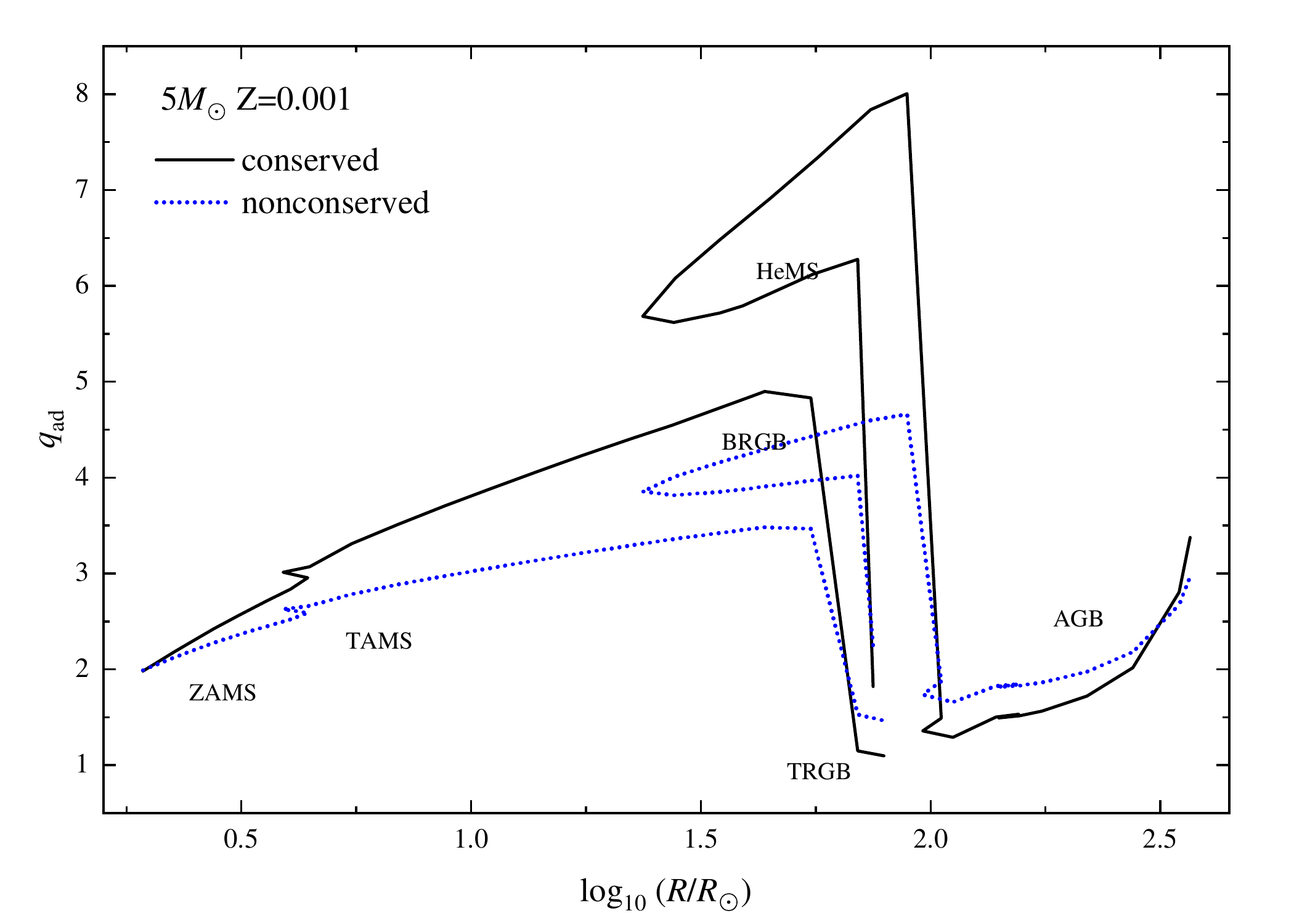}
	\caption{Critical mass ratios $q_\mathrm{ad}$ as functions of stellar radius for $5\,M_\odot$ stars. The solid (black) and dotted (blue) lines correspond to models of fully conservative and nonconservative mass transfer.}
	\label{fig-5m-qad}
\end{figure}

\begin{figure}[htb!]
	\centering
	\includegraphics[scale=0.25]{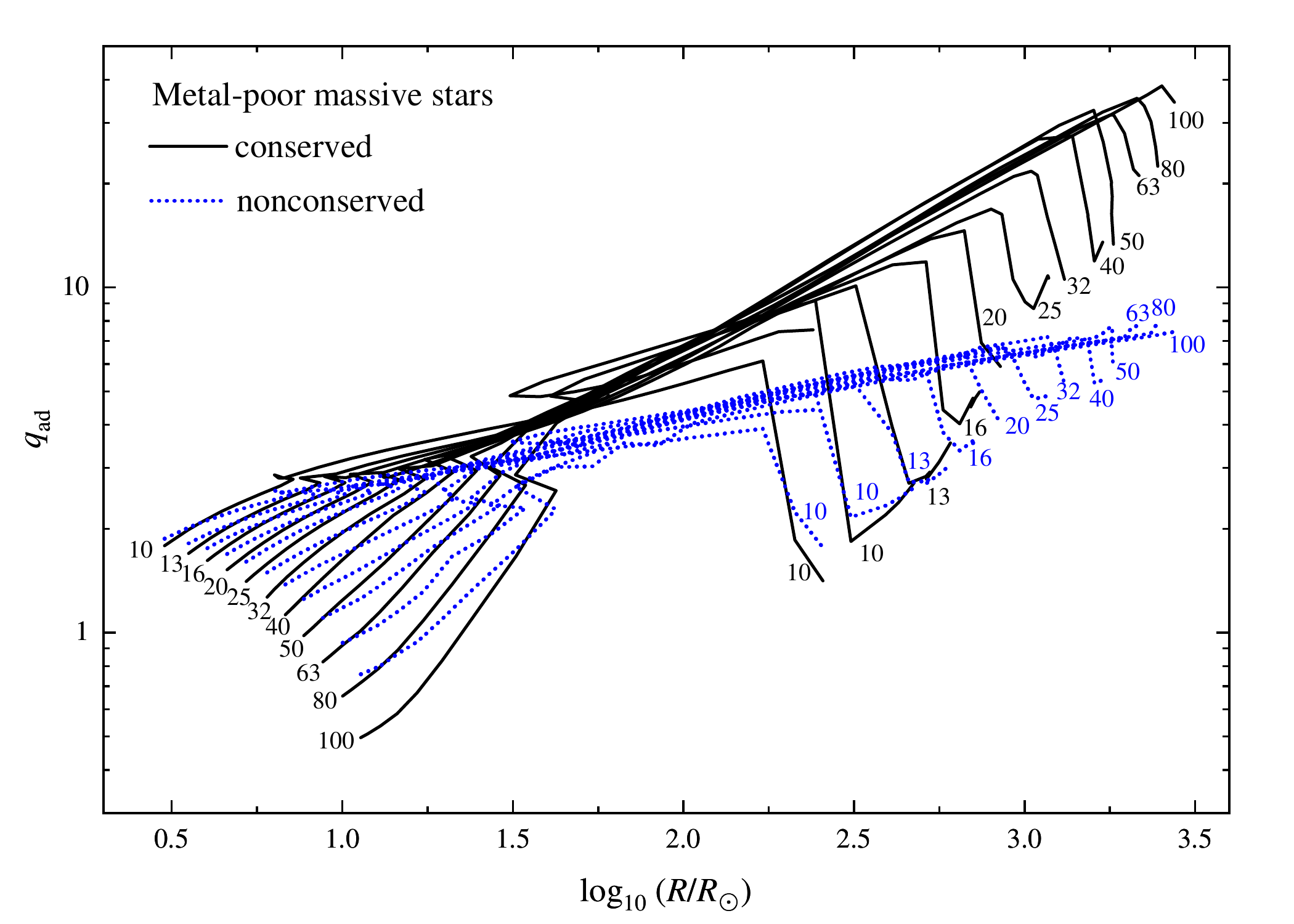}
	\caption{Critical mass ratios $q_\mathrm{ad}$ as functions of stellar radius for metal-poor massive stars. The masses in units of $M_\odot$ are labeled for each track. The solid (black) and dotted (blue) lines correspond to models of fully conservative and nonconservative mass transfer.}
	\label{fig-massive-qad}
\end{figure}

We plot the critical mass ratios $q_\mathrm{ad}$ of metal-poor intermediate-mass $5\,M_\odot$ and massive from $10$ to $100\,M_\odot$ stars in Figures\,\ref{fig-5m-qad} and \ref{fig-massive-qad}. During core-helium burning, the critical mass ratios increase dramatically because a deep convective envelope is replaced by a thick radiative envelope. Massive donors show similar trendsto the $20\,M_\odot$ stars. The critical mass ratio $q_\mathrm{ad}$ increases evolution of the donor until a convective envelope appears (Figure\,\ref{fig-massive-qad}).

\section{High Mass X-ray Binaries}
\label{sec-hmxb}

A high-mass X-ray binary (HMXB) is generally defined as a binary star system consisting of a neutron star (NS) or a BH accreting matter from a high-mass ($M \gtrsim 10 M_\odot$) stellar companion \citep{1972NPhS..239...67V,1973NInfo..27...70T}. Over 400 HMXBs have been found in the Milk Way (\citealt{2023A&A...671A.149F}), the Large (\citealt{2016MNRAS.459..528A}) and Small (\citealt{2016A&A...586A..81H}) Magellanic Clouds, and nearby galaxies (\citealt{2021ApJ...907...17L}). The donor stars in HMXBs are dominated by supergiant (Sg) O/B stars or Be stars (\citealt{2023hxga.book..143F}), and there are a few HMXBs with Wolf-Reyet donors or near naked helium stars. Other HMXBs include gamma-ray binaries that are dominated by gamma rays above 1Mev rather than X-rays and ultra-luminous , $L_\mathrm{X} > 1 \times 10^{39}\, \mathrm{erg\,s^{-1}}$, sources. HMXBs are important in modern astrophysics, such as first generations of HMXBs may provide X-ray heating to intergalactic gas and impact the onset and duration of Epoch of Reionization \citep{2014MNRAS.440.3778J}, understanding the formation and evolution of HMXBs provides important insights into one of the primary formation channels of gravitational wave sources \citep{2023hxga.book..143F}. 

Recently, updated catalogue of HMXBs have been provided by \citet{2023A&A...677A.134N} and \citet{2023A&A...671A.149F}. New HMXBs, including ultra-luminous X-ray sources \citep{2020AA...642A.174M}, have been found since the widely used catalogue of \citet{2006A&A...455.1165L}. From the observed HMXBs, we select 82 objects with known orbital information. We gather their orbital periods $P_\mathrm{orb}$, compact companion masses $M_\mathrm{X}$, donor masses $M_\mathrm{d}$, peak X-ray luminosities $L_\mathrm{X}$, and eccentricities $e$ along with other useful information in Table\,\ref{tab-hmxbs}. 

\begin{figure}[ht!]
	\centering
	\includegraphics[scale=0.28]{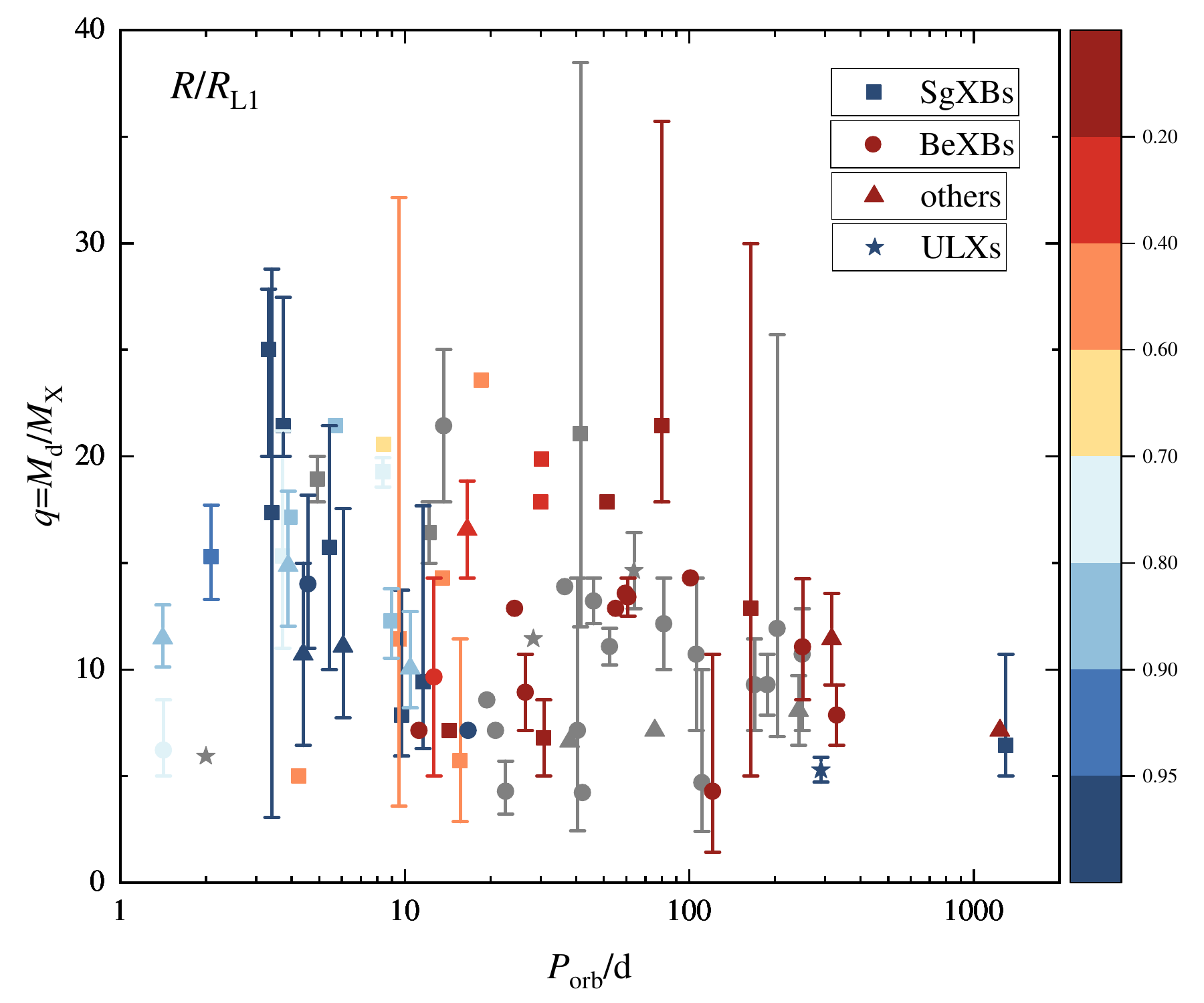}
	\caption{The ratio of donor radius $R$ to Roche-lobe radius $R_\mathrm{L}$ of 82 HMXBs in Table\,\ref{tab-hmxbs} on a mass ratio against the orbital period diagram. Symbols represent the types of HMXBs in the Table. Colours (gray means unknown) show the ratio of donor radius to its Roche-lobe radius. }
	\label{fig-rr}
\end{figure}

\begin{figure}[b!]
	\centering
	\includegraphics[scale=0.28]{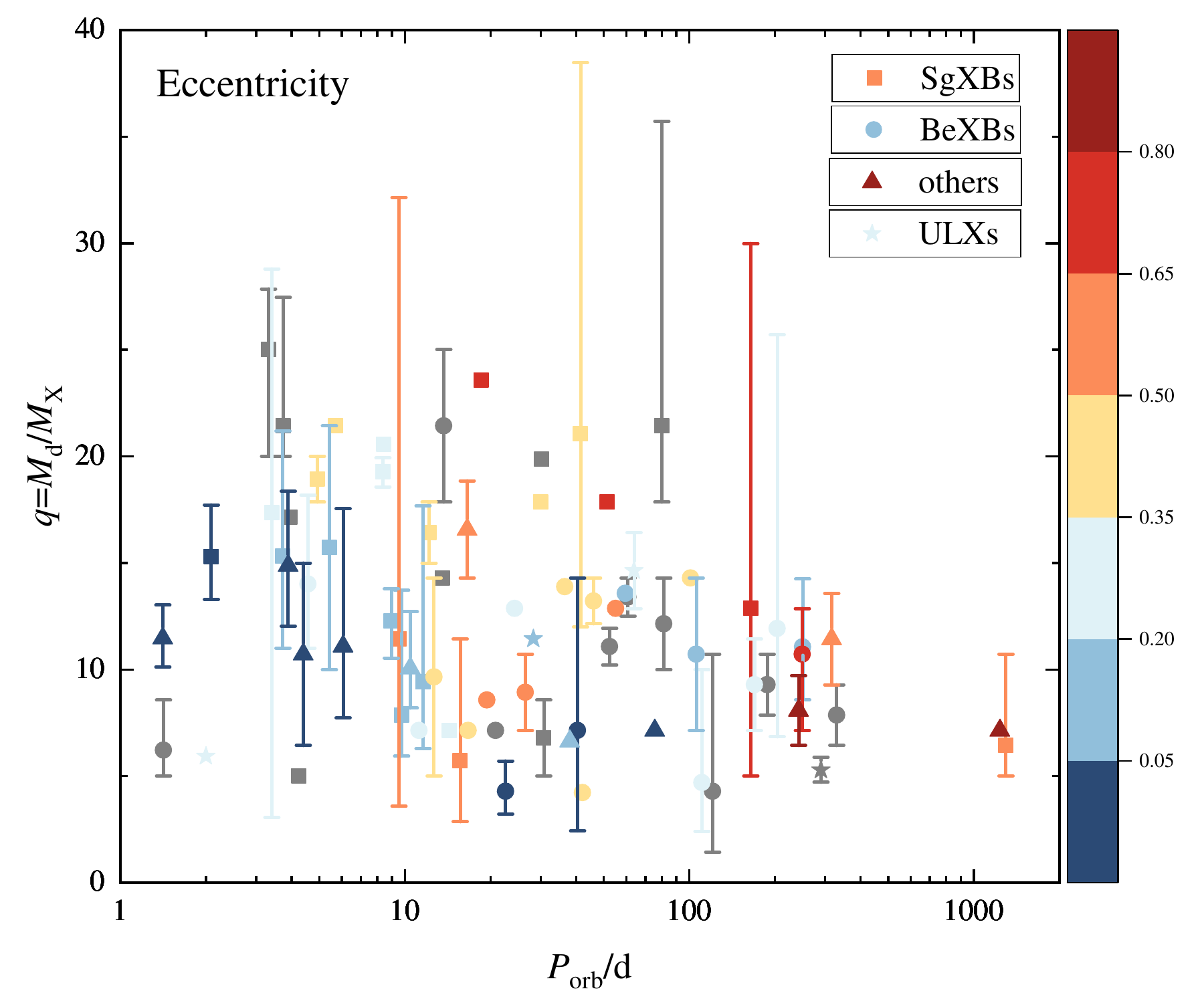}
	\caption{The eccentricities of HMXBs on a mass ratio orbital period diagram. Again symbols show the HMXB types.}
	\label{fig-ecc}
\end{figure}

We further select HMXBs undergoing stable mass transfer in circular orbits to compare our theoretical results. Unlike low and intermediate-mass X-ray binaries, most HMXBs are wind mass-transferring systems. So, the mass ratio $q = M_\mathrm{d}/M_\mathrm{X}$ of HMXBs can be large, over twenty (Figure\,\ref{fig-rr} and \ref{fig-ecc}). These are much larger than the mass ratios of intermediate-mass X-ray binaries, which reach $q_\mathrm{max} \approx 6$ (Figure 17 and Table 3 of \citealt{2023ApJ...945....7G}) that are undergoing stable Roche lobe overflow (RLOF). From Figure\,\ref{fig-rr}, we see that most of the relatively long orbital-period ($P_\mathrm{orb} > 10\,\mathrm{d}$) HMXBs have low ratios of donor radius to Roche-lobe radius. On the other hand at relatively short orbital periods ($P_\mathrm{orb} < 10\,\mathrm{d}$) most HMXBs are close to or undergoing stable RLOF.

Similarly, we find that relatively long orbital-period ($P_\mathrm{orb} > 10\,\mathrm{d}$) HMXBs can have a moderately ($e > 0.35$) to highly eccentric orbits (Figure\,\ref{fig-ecc}). On the contrary, HMXBs in short period ($P_\mathrm{orb} < 10\,\mathrm{d}$) are generally in low ($e < 0.35$) eccentricity orbits. When the progenitor of the accreting NS or BH explordes as a supernova, there is likely a natal kick caused by spatially asymmetric mass loss or neutrino emission \citep{1987IAUS..125..255W}. Such natal kicks can change the eccentricity and the final separation and even unbind otherwise runaway binary \citep{1975A&A....39...61F,1983ApJ...267..322H}. Here, we select HMXBs with a low eccentricity ($e < 0.35$) and a high filling degree ($R/R_\mathrm{L} > 0.95$) on the demands that they are closest to stable RLOF. 

\begin{longrotatetable}
\begin{deluxetable*}{llrlllllrrrl}
	\tabletypesize{\scriptsize}
	%\tablewidth{0pt}
	\tablecolumns{12}
	\rotate
	\tablecaption{Observed High-mass X-ray Binaries with known masses and orbital periods
	\label{tab-hmxbs}}
	\tablehead{
		 \colhead{Name} & \colhead{XrayType} & \colhead{$P_\mathrm{orb}$} 
		&\colhead{Peak $L_\mathrm{X}$}
		&\colhead{$q=M_\mathrm{d}/M_\mathrm{X}$} &\colhead{$M_\mathrm{X}$} 
		&\colhead{SpType*} & \colhead{$M_\mathrm{d}$} & \colhead{$e$} 
		&\colhead{$R_\mathrm{d}$} & \colhead{$R_\mathrm{L}$}
		&\colhead{References} \\
		\colhead{ } & \colhead{ } & \colhead{(d)} & \colhead{(${\rm erg \,s^{-1}}$) }
		&\colhead{ } &\colhead{($M_\odot$)} & \colhead{ } & \colhead{($M_\odot$)} & \colhead{ } 
		&\colhead{($R_\odot$)} & \colhead{($R_\odot$)}& \colhead{ } 
	}

 \startdata
 \multicolumn{12}{c}{SgXBs} \\
 \hline
4U 1954+31	&	SG	&	1296.64	&	\nodata	&$	6.43	^{+	4.29	}_{-	1.43	}$	&$	1.40	^{		}_{		}$	&	M4I	&$	9.00	^{+	6.00	}_{-	2.00	}$	&	0.500	&	586.0	&	591.9	&	1	\\
IGR J11215-5952	&	SG,XT	&	164.60	&	5.10E+36	&$	12.86	^{+	17.14	}_{-	7.86	}$	&$	1.40	^{		}_{		}$	&	B0.5Ia	&$	18.00	^{+	24.00	}_{-	11.00	}$	&	0.800	&	28.0	&	202.9	&	2--4	\\
IGR J16318-4848	&	RS,SG	&	80.00	&	4.00E+36	&$	21.43	^{+	14.29	}_{-	3.57	}$	&$	1.40	^{		}_{		}$	&	B0-5sgBe	&$	30.00	^{+	20.00	}_{-	5.00	}$	&	\nodata	&	20.4	&	156.3	&	5, 6	\\
IGR J17391-3021	&	SG,XT	&	51.47	&	1.00E+36	&$	17.86	^{		}_{		}$	&$	1.40	^{		}_{		}$	&	O8Iab(f)	&$	25.00	^{		}_{		}$	&	$<0.800$	&	18.8	&	107.8	&	2, 7, 8	\\
GX 301-2	&	CL,SG,XP,XT	&	41.50	&	3.00E+37	&$	21.05	^{+	17.41	}_{-	9.05	}$	&$	1.90	^{+	0.60	}_{-	0.60	}$	&	B1.5Ia	&$	40.00	^{+	10.00	}_{-	10.00	}$	&	0.470	&	\nodata	&	110.9	&	9--12	\\
IGR J05007-7047	&	SG	&	30.77	&	9.00E+36	&$	6.79	^{+	1.79	}_{-	1.79	}$	&$	1.40	^{		}_{		}$	&	B2III	&$	9.50	^{+	2.50	}_{-	2.50	}$	&	\nodata	&	4.8	&	50.1	&	13	\\
IGR J16465-4507	&	SG,XP,XT	&	30.32	&	6.80E+36	&$	19.86	^{		}_{		}$	&$	1.40	^{		}_{		}$	&	B0.5-1Ib	&$	27.80	^{		}_{		}$	&	\nodata	&	22.1	&	79.3	&	2, 14	\\
SAX J1818.6-1703	&	SG,XT	&	30.00	&	8.00E+35	&$	17.86	^{		}_{		}$	&$	1.40	^{		}_{		}$	&	B0.5Iab	&$	25.00	^{		}_{		}$	&	0.350	&	30.0	&	75.2	&	15, 16	\\
IGR J18483-0311	&	SG,XT	&	18.55	&	7.80E+36	&$	23.57	^{		}_{		}$	&$	1.40	^{		}_{		}$	&	B0.5-1Iab	&$	33.00	^{		}_{		}$	&	0.40-0.68	&	33.8	&	61.4	&	2, 4, 17	\\
IGR J00370+6122	&	SG	&	15.66	&	3.00E+36	&$	5.71	^{+	5.71	}_{-	2.86	}$	&$	1.40	^{		}_{		}$	&	BN0.7Ib	&$	8.00	^{+	8.00	}_{-	4.00	}$	&	0.560	&	17.0	&	29.6	&	3, 18, 19	\\
1E 1145.1-6141	&	RS,SG,XP	&	14.37	&	4.70E+37	&$	7.14	^{		}_{		}$	&$	1.40	^{		}_{		}$	&	B2Iae	&$	10.00	^{		}_{		}$	&	0.200	&	5.0	&	30.8	&	4, 20	\\
IGR J19140+0951	&	SG	&	13.56	&	3.00E+35	&$	14.29	^{		}_{		}$	&$	1.40	^{		}_{		}$	&	B0.5Ia	&$	20.00	^{		}_{		}$	&	\nodata	&	20.0	&	40.2	&	2, 21	\\
IGR J17544-2619	&	SG,XT	&	12.17	&	5.30E+36	&$	16.43	^{+	1.43	}_{-	1.43	}$	&$	1.40	^{		}_{		}$	&	O9Ib	&$	23.00	^{+	2.00	}_{-	2.00	}$	&	0.440	&	\nodata	&	39.8	&	4, 22--25	\\
2S 0114+650	&	CL,SG,XP	&	11.60	&	1.20E+37	&$	9.41	^{+	8.27	}_{-	3.12	}$	&$	1.70	^{+	0.10	}_{-	0.50	}$	&	B1Iae	&$	16.00	^{+	5.00	}_{-	5.00	}$	&	0.160	&	37.0	&	32.2	&	2, 4, 26, 27	\\
IGR J17252-3616	&	EB,SG,XP	&	9.74	&	1.60E+37	&$	7.85	^{+	5.85	}_{-	1.92	}$	&$	1.91	^{+	0.45	}_{-	0.45	}$	&	B0-B1Ia	&$	15.00	^{+	5.00	}_{-	1.00	}$	&	$<0.190$	&	29.0	&	27.5	&	2, 28	\\
4U 2206+54	&	BE,CL,QPO,SG,XP	&	9.57	&	5.50E+35	&$	11.43	^{		}_{		}$	&$	1.40	^{		}_{		}$	&	O9.5Vep	&$	16.00	^{		}_{		}$	&	0.150	&	7.3	&	28.9	&	3, 4, 29	\\
IGR J08408-4503	&	SG,XT	&	9.54	&	6.00E+36	&$	11.43	^{+	20.71	}_{-	7.86	}$	&$	1.40	^{		}_{		}$	&	O8.5Ib-II(f)p	&$	16.00	^{+	29.00	}_{-	11.00	}$	&	0.630	&	17.0	&	28.9	&	3, 30	\\
Vela X-1	&	CL,EB,RS,SG,XP	&	8.96	&	1.00E+37	&$	12.26	^{+	1.52	}_{-	1.73	}$	&$	2.12	^{+	0.16	}_{-	0.16	}$	&	B0.5Ib	&$	26.00	^{+	1.00	}_{-	2.00	}$	&	0.090	&	29.0	&	32.8	&	4, 28, 31, 32	\\
IGR J17354-3255	&	SG,XT	&	8.45	&	2.60E+36	&$	20.57	^{		}_{		}$	&$	1.40	^{		}_{		}$	&	O9.5Iab	&$	28.80	^{		}_{		}$	&	0.200	&	22.4	&	34.3	&	2, 33	\\
4U 1907+09	&	CL,HT,SG,XP,XT	&	8.38	&	4.80E+36	&$	19.29	^{+	0.64	}_{-	0.71	}$	&$	1.40	^{		}_{		}$	&	O8-9Ia	&$	27.00	^{+	0.90	}_{-	1.00	}$	&	0.280	&	26.2	&	33.2	&	4, 34, 35	\\
IGR J18450-0435	&	SG,XT	&	5.72	&	8.00E+36	&$	21.43	^{+	0.00	}_{-	0.00	}$	&$	1.40	^{		}_{		}$	&	O9.5I	&$	30.00	^{+		}_{-		}$	&	$<0.370$	&	23.0	&	26.9	&	36	\\
IGR J18214-1318	&	SG	&	5.42	&	5.00E+36	&$	15.71	^{+	5.71	}_{-	5.71	}$	&$	1.40	^{		}_{		}$	&	B0V-O9I	&$	22.00	^{+	8.00	}_{-	8.00	}$	&	0.170	&	22.0	&	22.8	&	37, 38	\\
IGR J17544-2619	&	CL,SG,XT	&	4.93	&	3.00E+38	&$	18.93	^{+	1.07	}_{-	1.07	}$	&$	1.40	^{		}_{		}$	&	O9Ib	&$	26.50	^{+	1.50	}_{-	1.50	}$	&	$<0.400$	&	\nodata	&	23.1	&	22, 39	\\
IGR J18027-2016	&	SG,CL,EB,QPO,XP,XT	&	4.57	&	1.30E+37	&$	14.01	^{+	4.17	}_{-	3.02	}$	&$	1.57	^{+	0.25	}_{-	0.25	}$	&	B1Ib	&$	22.00	^{+	2.00	}_{-	2.00	}$	&	0.200	&	20.0	&	20.1	&	2,4, 28, 40	\\
IGR J16393-4643	&	CL,SG,XB,XP,EB	&	4.24	&	\nodata	&$	5.00	^{		}_{		}$	&$	1.40	^{		}_{		}$	&	BIV-V(OB)	&$	7.00	^{		}_{		}$	&	\nodata	&	6.3	&	11.7	&	40, 41	\\
IGR J16195-4945	&	SG	&	3.95	&	\nodata	&$	17.14	^{		}_{		}$	&$	1.40	^{		}_{		}$	&	ON9.7Iab	&$	24.00	^{		}_{		}$	&	\nodata	&	17.0	&	19.1	&	42	\\
IGR J16418-4532	&	SG,XT,EB	&	3.74	&	2.10E+37	&$	21.43	^{+	6.03	}_{-	1.43	}$	&$	1.40	^{		}_{		}$	&	O7.5Ia	&$	30.00	^{+	8.44	}_{-	2.00	}$	&	\nodata	&	21.7	&	20.3	&	2, 4, 41	\\
4U 1538-52	&	CL,EB,SG,XP	&	3.73	&	9.00E+36	&$	15.31	^{+	5.87	}_{-	4.31	}$	&$	1.30	^{+	0.20	}_{-	0.20	}$	&	B0Iab	&$	19.90	^{+	3.40	}_{-	3.40	}$	&	0.180	&	13.0	&	17.1	&	11, 28, 43	\\
4U 1700-377	&	EB,RS,SG	&	3.41	&	2.20E+36	&$	17.35	^{+	11.46	}_{-	14.29	}$	&$	1.96	^{+	0.19	}_{-	0.19	}$	&	O6Iafcp	&$	34.00	^{+	17.00	}_{-	28.00	}$	&	$<0.220$	&	19.0	&	19.5	&	3, 28	\\
IGR J16479-4514	&	SG,XT,EB	&	3.32	&	1.09E+34	&$	25.00	^{+	2.86	}_{-	5.00	}$	&$	1.40	^{		}_{		}$	&	O7I	&$	35.00	^{+	4.00	}_{-	7.00	}$	&	\nodata	&	20.5	&	20.0	&	41	\\
Cen X-3	&	CL,EB,QPO,SG,XP	&	2.09	&	5.40E+37	&$	15.29	^{+	2.44	}_{-	2.00	}$	&$	1.57	^{+	0.16	}_{-	0.16	}$	&	O6.5II-III	&$	24.00	^{+	1.00	}_{-	1.00	}$	&	$<0.002$	&	11.4	&	12.4	&	11, 28,	\\
\hline
 \multicolumn{12}{c}{BeXBs} \\
\hline
RX J0146.9+6121	&	BE	&	330.00	&	1.10E+35	&$	7.86	^{+	1.43	}_{-	1.43	}$	&$	1.40	^{		}_{		}$	&	B1III-Ve	&$	11.00	^{+	2.00	}_{-	2.00	}$	&	\nodata	&	7.0	&	259.8	&	44	\\
4U 0352+30	&	BE,CL,XP	&	250.30	&	6.30E+34	&$	11.06	^{+	3.19	}_{-	2.48	}$	&$	1.40	^{		}_{		}$	&	B0Ve	&$	15.49	^{+	4.46	}_{-	3.47	}$	&	0.110	&	6.4	&	251.2	&	2, 4, 45	\\
GRO J1008-57	&	BE,CL,HT,XP,XT	&	249.50	&	1.00E+38	&$	10.71	^{+	2.14	}_{-	3.57	}$	&$	1.40	^{		}_{		}$	&	B0IIIVe	&$	15.00	^{+	3.00	}_{-	5.00	}$	&	0.680	&	\nodata	&	247.2	&	11, 46, 47	\\
gam Cas	&	BE,Gcas,RS	&	203.59	&	\nodata	&$	11.92	^{+	13.79	}_{-	5.08	}$	&$	1.30	^{+	0.60	}_{-	0.60	}$	&	B0.5IVpe	&$	15.50	^{+	2.50	}_{-	2.50	}$	&	0.260	&	\nodata	&	220.7	&	48	\\
4U 1145-619	&	BE,XP,XT	&	187.50	&	1.90E+35	&$	9.29	^{+	1.43	}_{-	1.43	}$	&$	1.40	^{		}_{		}$	&	B0.2III	&$	13.00	^{+	2.00	}_{-	2.00	}$	&	\nodata	&	\nodata	&	191.9	&	4, 49--51	\\
XTE J1946+274	&	BE,CL,XP,XT	&	169.20	&	5.00E+37	&$	9.29	^{+	2.14	}_{-	2.14	}$	&$	1.40	^{		}_{		}$	&	B0-1IV-Ve	&$	13.00	^{+	3.00	}_{-	3.00	}$	&	0.330	&	\nodata	&	179.2	&	52	\\
IGR J11305-6256	&	BE	&	120.83	&	1.00E+35	&$	4.29	^{+	6.43	}_{-	2.86	}$	&$	1.40	^{		}_{		}$	&	B0IIIne	&$	6.00	^{+	9.00	}_{-	4.00	}$	&	\nodata	&	6.0	&	101.7	&	3, 53	\\
1A 0535+262	&	BE,CL,HT,QPO,RS,XP,XT	&	110.60	&	4.90E+37	&$	4.69	^{+	5.31	}_{-	2.27	}$	&$	1.60	^{+	0.60	}_{-	0.60	}$	&	O9.7IIIe	&$	7.50	^{+	2.50	}_{-	2.50	}$	&	0.350	&	\nodata	&	104.3	&	11, 51	\\
GS 0834-430	&	BE,XP,XT	&	105.80	&	1.10E+37	&$	10.71	^{+	3.57	}_{-	3.57	}$	&$	1.40	^{		}_{		}$	&	B0-2III-Ve	&$	15.00	^{+	5.00	}_{-	5.00	}$	&	0.120	&	\nodata	&	139.5	&	50, 54	\\
MXB 0656-072	&	BE,CL,QPO,XP,XT	&	101.00	&	6.60E+36	&$	14.29	^{		}_{		}$	&$	1.40	^{		}_{		}$	&	O9.7Ve	&$	20.00	^{		}_{		}$	&	0.400	&	15.0	&	153.4	&	55	\\
LS 992	&	BE,XP,XT	&	81.30	&	2.30E+36	&$	12.14	^{+	2.14	}_{-	2.14	}$	&$	1.40	^{		}_{		}$	&	B0.2IVe	&$	17.00	^{+	3.00	}_{-	3.00	}$	&	\nodata	&	\nodata	&	123.7	&	4, 50, 56	\\
LS 1698	&	BE,XP,XT	&	60.90	&	4.50E+35	&$	13.39	^{+	0.89	}_{-	0.89	}$	&$	1.40	^{		}_{		}$	&	B0V-IIIe	&$	18.75	^{+	1.25	}_{-	1.25	}$	&	\nodata	&	11.2	&	106.4	&	57, 58	\\
SGR 0755-2933	&	BE	&	59.52	&	1.00E+34	&$	13.57	^{		}_{		}$	&$	1.40	^{		}_{		}$	&	B0Ve	&$	19.00	^{		}_{		}$	&	0.060	&	12.6	&	105.4	&	4, 59	\\
GRO J2058+42	&	BE,CL,XP,XT	&	55.03	&	5.60E+37	&$	12.86	^{		}_{		}$	&$	1.40	^{		}_{		}$	&	O9.5-B0IV-Ve	&$	18.00	^{		}_{		}$	&	0.600	&	10.0	&	97.7	&	60	\\
IGR J11435-6109	&	BE,XP,XT	&	52.46	&	2.10E+36	&$	11.07	^{+	0.86	}_{-	0.86	}$	&$	1.40	^{		}_{		}$	&	B0.5Ve	&$	15.50	^{+	1.20	}_{-	1.20	}$	&	\nodata	&	\nodata	&	88.7	&	4, 61	\\
EXO 2030+375	&	BE,CL,XP,XT	&	46.02	&	1.00E+38	&$	13.21	^{+	1.07	}_{-	1.07	}$	&$	1.40	^{		}_{		}$	&	B0Ve	&$	18.50	^{+	1.50	}_{-	1.50	}$	&	0.410	&	\nodata	&	87.8	&	62, 63	\\
2S 1417-624	&	BE,XP,XT	&	42.12	&	8.00E+36	&$	4.21	^{		}_{		}$	&$	1.40	^{		}_{		}$	&	B1Ve	&$	>5.9	^{		}_{		}$	&	0.446	&	\nodata	&	50.0	&	50, 64	\\
KS 1947+300	&	BE,CL,HT,XP,XT	&	40.42	&	2.00E+37	&$	7.14	^{+	7.14	}_{-	4.71	}$	&$	1.40	^{		}_{		}$	&	B0Ve	&$	10.00	^{+	10.00	}_{-	6.60	}$	&	0.030	&	\nodata	&	61.4	&	65, 66	\\
V0332+53	&	BE,CL,HT,QPO,XP,XT	&	36.50	&	3.75E+38	&$	13.89	^{		}_{		}$	&$	1.44	^{		}_{		}$	&	O8-9Ve	&$	>20	^{		}_{		}$	&	0.417	&	\nodata	&	77.6	&	11, 67	\\
LS I +61 303	&	BE,GP,RS	&	26.50	&	\nodata	&$	8.93	^{+	1.79	}_{-	1.79	}$	&$	1.40	^{		}_{		}$	&	B0Ve	&$	12.50	^{+	2.50	}_{-	2.50	}$	&	0.540	&	7.0	&	51.2	&	68	\\
4U 0115+63	&	BE,CL,HT,QPO,XB,XP,XT	&	24.30	&	1.02E+38	&$	12.86	^{		}_{		}$	&$	1.40	^{		}_{		}$	&	B0.2Ve	&$	18.00	^{		}_{		}$	&	0.340	&	8.0	&	56.7	&	69, 70	\\
4U 1901+03	&	BE,CL,XP,XT	&	22.58	&	1.10E+38	&$	4.29	^{+	1.43	}_{-	1.07	}$	&$	1.40	^{		}_{		}$	&	B8-9 IVe	&$	6.00	^{+	2.00	}_{-	1.50	}$	&	0.036	&	\nodata	&	33.2	&	71	\\
Cep X-4	&	BE,CL,HT,XP,XT	&	20.85	&	1.40E+36	&$	7.14	^{		}_{		}$	&$	1.40	^{		}_{		}$	&	B1-B2Ve	&$	10.00	^{		}_{		}$	&	\nodata	&	\nodata	&	39.5	&	11, 72	\\
XTE J0421+560	&	BE,SG,XT	&	19.41	&	3.00E+38	&$	8.57	^{		}_{		}$	&$	1.40	^{		}_{		}$	&	B1/2I[e]	&$	>12	^{		}_{		}$	&	0.620	&	\nodata	&	40.8	&	73, 74	\\
Cir X-1	&	BE?	&	16.68	&	1.00E+38	&$	7.14	^{		}_{		}$	&$	1.40	^{		}_{		}$	&	B5-A0I	&$	10.00	^{		}_{		}$	&	0.450	&	40.0	&	34.0	&	75, 76	\\
IGR J14059-6116	&	GP,BE	&	13.71	&	5.60E+33	&$	21.43	^{+	3.57	}_{-	3.57	}$	&$	1.40	^{		}_{		}$	&	O6.5III	&$	30.00	^{+	5.00	}_{-	5.00	}$	&	\nodata	&	\nodata	&	48.2	&	77, 78	\\
SAX J2103.5+4545	&	BE,XP,XT	&	12.68	&	8.40E+36	&$	9.64	^{+	4.64	}_{-	4.64	}$	&$	1.40	^{		}_{		}$	&	B0Ve	&$	13.50	^{+	6.50	}_{-	6.50	}$	&	0.400	&	8.0	&	32.4	&	4, 79, 80	\\
PSR J0635+0533	&	BE,XP	&	11.20	&	2.76E+33	&$	7.14	^{		}_{		}$	&$	1.40	^{		}_{		}$	&	B1IIIe-B2Ve	&$	10.00	^{		}_{		}$	&	0.290	&	\nodata	&	26.1	&	4, 81	\\
RX J0050.7-7316	&	BE	&	1.42	&	1.00E+36	&$	6.21	^{+	2.36	}_{-	1.21	}$	&$	1.40	^{		}_{		}$	&	B	&$	8.70	^{+	3.30	}_{-	1.70	}$	&	\nodata	&	4.3	&	6.2	&	41, 82	\\
\hline
 \multicolumn{12}{c}{Other XBs} \\
\hline
PSR B1259-63	&	GP	&	1236.72	&	8.00E+35	&$	7.14	^{		}_{		}$	&$	1.40	^{		}_{		}$	&	B2e/O9.5Ve	&$	10.00	^{		}_{		}$	&	0.870	&	6.0	&	600.9	&	2, 83, 84	\\
HESS J0632+057	&	GP	&	316.80	&	\nodata	&$	11.43	^{+	2.14	}_{-	2.14	}$	&$	1.40	^{		}_{		}$	&	B0Vpe	&$	16.00	^{+	3.00	}_{-	3.00	}$	&	0.620	&	8.0	&	298.2	&	85, 86	\\
2S 1845-024	&	XP,XT	&	242.18	&	3.00E+37	&$	8.07	^{+	1.64	}_{-	1.64	}$	&$	1.40	^{		}_{		}$	&	OB	&$	11.30	^{+	2.30	}_{-	2.30	}$	&	0.880	&	\nodata	&	213.9	&	4, 87, 88	\\
XTE J1543-568	&	XP,XT	&	75.56	&	1.20E+37	&$	7.14	^{		}_{		}$	&$	1.40	^{		}_{		}$	&	Be?	&$	>10	^{		}_{		}$	&	0.030	&	\nodata	&	93.2	&	50, 89	\\
XTE J1859+083	&	XP,XT	&	37.97	&	1.00E+36	&$	6.64	^{		}_{		}$	&$	1.40	^{		}_{		}$	&	B0-2Ve	&$	>9.3	^{		}_{		}$	&	0.127	&	\nodata	&	57.1	&	90--92	\\
1FGL J1018.6-5856	&	GP	&	16.55	&	\nodata	&$	16.57	^{+	2.29	}_{-	2.29	}$	&$	1.40	^{		}_{		}$	&	O6V((f))	&$	23.20	^{+	3.20	}_{-	3.20	}$	&	0.530	&	10.0	&	49.0	&	51, 93	\\
OAO 1657-415	&	CL,EB,XP	&	10.45	&	2.00E+37	&$	10.06	^{+	2.65	}_{-	1.87	}$	&$	1.74	^{+	0.30	}_{-	0.30	}$	&	B0-6sg	&$	17.50	^{+	0.80	}_{-	0.80	}$	&	0.103	&	25.0	&	31.2	&	4, 28, 94	\\
XTE J1855-026	&	EB,XP,XT	&	6.07	&	1.90E+37	&$	11.09	^{+	6.45	}_{-	3.35	}$	&$	1.80	^{	0.59	}_{-	0.58	}$	&	B0Iaep	&$	19.90	^{+	1.50	}_{-	1.40	}$	&	$<0.040$	&	22.0	&	22.9	&	4, 28, 41	\\
4U 1909+07	&	CL,XP	&	4.40	&	3.50E+36	&$	10.71	^{+	4.29	}_{-	4.29	}$	&$	1.40	^{		}_{		}$	&	O7.5-9.5If	&$	15.00	^{+	6.00	}_{-	6.00	}$	&	0.021	&	16.0	&	16.7	&	4, 95	\\
SMC X-1	&	EB	&	3.89	&	\nodata	&$	14.88	^{+	3.47	}_{-	2.85	}$	&$	1.21	^{+	0.12	}_{-	0.12	}$	&	B0sg	&$	18.00	^{+	2.00	}_{-	2.00	}$	&	0.001	&	15.0	&	17.0	&	28	\\
LMC X-4	&	EB	&	1.41	&	3.60E+38	&$	11.46	^{+	1.55	}_{-	1.34	}$	&$	1.57	^{+	0.11	}_{-	0.11	}$	&	O8III	&$	18.00	^{+	1.00	}_{-	1.00	}$	&	0.006	&	7.4	&	8.4	&	28, 96	\\
\hline
 \multicolumn{12}{c}{NS-ULXs} \\
\hline
NGC 300 ULX-1	&	ULX	&	290.00	&	5.00E+39	&$	5.29	^{+	0.59	}_{-	0.59	}$	&$	1.70	^{+	0.30	}_{-	0.30	}$	&	\nodata	&$	9	^{+	1	}_{-	1	}$	&	\nodata	&	310.0	&	213.5	&	97, 98	\\
NGC 7793 P13	&	ULX	&	64.00	&	5.00E+39	&$	14.64	^{+	1.79	}_{-	1.79	}$	&$	1.40	^{		}_{		}$	&	B9I	&$	20.50	^{+	2.50	}_{-	2.50	}$	&	0.325	&	\nodata	&	114.4	&	99--101	\\
Swift J0243.6+6124	&	BE,CL,RS,ULX	&	28.30	&	2.00E+39	&$	11.43	^{		}_{		}$	&$	1.40	^{		}_{		}$	&	O9.5Ve	&$	16.00	^{		}_{		}$	&	0.100	&	\nodata	&	59.6	&	4, 102, 103	\\
M51 ULX-7	&	ULX	&	2.00	&	1.00E+40	&$	5.93	^{		}_{		}$	&$	1.40	^{		}_{		}$	&	\nodata	&$	>8.3	^{		}_{		}$	&	0.220	&	\nodata	&	7.6	&	104, 105	\\
\hline
 \multicolumn{12}{c}{BHXBs} \\
\hline
MWC 656	&	BE,BH	&	60.37	&	1.00E+32	&$	2.44	^{+	0.56	}_{-	0.56	}$	&$	5.33	^{+	1.57	}_{-	1.53	}$	&	B1.5-B2III	&$	13.00	^{+	3.00	}_{-	3.00	}$	&	0.100	&	10.0	&	78.1	&	106	\\
SS 433	&	BE,BH,EB,MQ,RS	&	13.10	&	2.00E+37	&$	2.69	^{+	0.14	}_{-	0.14	}$	&$	4.20	^{+	0.40	}_{-	0.40	}$	&	A3-7I	&$	11.30	^{+	0.60	}_{-	0.60	}$	&	\nodata	&	27.2	&	27.2	&	107	\\
Cyg X-1	&	BH,MQ,RS,SG,US	&	5.60	&	1.40E+37	&$	1.92	^{+	0.36	}_{-	0.33	}$	&$	21.20	^{+	2.20	}_{-	2.20	}$	&	O9.7Iabpvar	&$	40.60	^{+	7.70	}_{-	7.10	}$	&	0.019	&	22.3	&	22.9	&	108	\\
HD 96670	&	BE,BH	&	5.28	&	2.40E+34	&$	3.66	^{+	0.84	}_{-	0.58	}$	&$	6.20	^{+	0.90	}_{-	0.70	}$	&	O8.5f(n)p	&$	22.70	^{+	5.20	}_{-	3.60	}$	&	0.120	&	17.1	&	19.3	&	109	\\
LS 5039	&	BH,GP,RS	&	3.91	&	6.00E+34	&$	6.19	^{+	0.92	}_{-	0.78	}$	&$	3.70	^{+	1.30	}_{-	1.00	}$	&	ON6V((f))z	&$	22.90	^{+	3.40	}_{-	2.90	}$	&	0.350	&	9.3	&	16.8	&	2, 110	\\
SAX J1819.3-2525	&	BH,XT	&	2.82	&	2.88E+39	&$	0.68	^{+	0.17	}_{-	0.11	}$	&$	9.61	^{+	2.08	}_{-	0.88	}$	&	B9III	&$	6.53	^{+	1.60	}_{-	1.03	}$	&	\nodata	&	7.5	&	7.3	&	111	\\
Cyg X-3	&	BH,MQ,RS	&	0.20	&	5.00E+38	&$	1.61	^{+	0.17	}_{-	0.17	}$	&$	7.20	^{		}_{		}$	&	WN4-6	&$	11.60	^{+	1.20	}_{-	1.20	}$	&	\nodata	&	8.0	&	1.6	&	112	\\
  \enddata
    \tablecomments{* Donor Spectral Type}
	\tablerefs{\\
1.\,\citet{2020ApJ...904..143H}, 2.\,\citet{2015AA...579A.111K}, 3.\,\citet{2020AA...634A..49H},  4.\,\citet{2023ApJS..268...21K}, 5.\,\citet{2012ApJ...751..150C}, 6.\,\citet{2020ApJ...894...86F}, 
7.\,\citet{2010MNRAS.409.1220D}, 8.\,\citet{2011ApJ...727...59B}, 9.\,\citet{1997ApJ...479..933K}, 
10.\,\citet{2006Msngr.126...27K}, 11.\,\citet{2019AA...622A..61S}, 12.\,\citet{2023ApJS..268...21K}, 
13.\,\citet{2011AA...529A..30D}, 14.\,\citet{2010MNRAS.405L..66L}, 15.\,\citet{2010AA...510A..61T},  16.\,\citet{2016MNRAS.456.4111B}, 17.\,\citet{2010MNRAS.401.1564R}, 18.\,\citet{2007AA...469.1063I},  19.\,\citet{2021PASJ...73.1389U}, 20.\,\citet{2008AA...479..533F},	21.\,\citet{2008MNRAS.389..301P}, 
22.\,\citet{2006AA...455..653P}, 23.\,\citet{2012AA...539A..21D}, 24.\,\citet{2013BCrAO.109...27N}, 25.\,\citet{2017AstL...43..664B}, 26.\,\citet{1996AA...311..879R}, 27.\,\citet{2008MNRAS.389..608F},  28.\,\citet{2015AA...577A.130F}, 29.\,\citet{2006AA...449..687R}, 30.\,\citet{2007ApJ...655L.101G}, 31.\,\citet{2021AA...652A..95K}, 32.\,\citet{2020MNRAS.492.5922L}, 33.\,\citet{2011MNRAS.417..573S}, 34.\,\citet{2005AA...436..661C}, 35.\,\citet{2013ApJ...771...96M}, 36.\,\citet{2013MNRAS.434.2182G}, 37.\,\citet{2009ApJ...698..502B}, 38.\,\citet{2020MNRAS.498.2750C}, 39.\,\citet{2015AA...576L...4R}, 40.\,\citet{2008AA...484..783C}, 41.\,\citet{2015ApJ...808..140C}, 42.\,\citet{2016MNRAS.456.2717C}, 43.\,\citet{1992MNRAS.256..631R}, 44.\,\citet{1997AA...322..183R}, 45.\,\citet{2001AA...380..615C}, 46.\,\citet{2007MNRAS.378.1427C}, 47.\,\citet{2014PASJ...66...59Y}, 48.\,\citet{2000AA...364L..85H}, 49.\,\citet{1997MNRAS.288..988S}, 50.\,\citet{2009ApJ...707..870B}, 51.\,\citet{2023MNRAS.525.1498Z}, 52.\,\citet{2003ApJ...584..996W}, 53.\,\citet{2006AA...449.1139M}, 54.\,\citet{1997ApJ...479..388W}, 55.\,\citet{2012ApJ...753...73Y}, 56.\,\citet{2001AA...367..266R}, 57.\,\citet{1997AA...323..853M}, 58.\,\citet{2013MNRAS.436L..74C}, 59.\,\citet{2023Natur.614...45R}, 60.\,\citet{2023AA...671A..48R}, 61.\,\citet{2022MNRAS.517.3034B}, 62.\,\citet{1999ApJ...512..313S}, 63.\,\citet{2023AA...675A..29M}, 64.\,\citet{1996AAS..120C.209F}, 65.\citet{2003AA...397..739N}, 66.\,\citet{2004ApJ...613.1164G}, 67.\,\citet{2010MNRAS.406.2663R}, 68.\,\citet{2005MNRAS.360.1105C}, 69.\,\citet{2001AA...369..108N}, 70.\,\citet{2013AA...551A...6M}, 71.\,\citet{2005ApJ...635.1217G}, 72.\,\citet{1991ApJ...366L..19K} 73.\,\citet{2006ASPC..355..305B}, 74.\,\citet{2019AA...622A..93B}, 75.\,\citet{2007MNRAS.374..999J}, 76.\,\citet{2020ApJ...891..150S}, 77.\,\citet{2019ApJ...884...93C}, 78.\,\citet{2023AA...680A..88S}, 79.\,\citet{2000ApJ...544L.129B}, 80.\,\citet{2004AA...421..673R}, 81.\,\citet{2000ApJ...542L..41K}, 82.\,\citet{2000MNRAS.311..169C}, 83.\,\citet{2008MNRAS.385.2279S}, 84.\,\citet{2011ApJ...732L..11N}, 85.\,\citet{2012MNRAS.421.1103C}, 86\,\citet{2018PASJ...70...61M}, 87.\,\citet{1999ApJ...517..449F}, 88.\,\citet{2022AA...657A..58N}, 89.\,\citet{2001ApJ...553L.165I}, 90.\,\citet{2009ApJ...695...30C}, 91.\,\citet{2016PhDT.......406B}, 92.\citet{2022MNRAS.509.5955S}, 93.\,\citet{2022MNRAS.515.1078V}, 94.\,\citet{2009AA...505..281M}, 95.\,\citet{2015AA...578A.107M}, 96.\,\citet{2023MNRAS.519.1764S}, 97.\,\citet{2019ApJ...883L..34H}, 98.\,\citet{2023NewAR..9601672K}, 99.\,\citet{2011AN....332..367M}, 100.\,\citet{2014Natur.514..198M}, 101.\,\citet{2020AA...642A.174M}, 102.\,\citet{2018AA...613A..19D}, 103.\,\citet{2020AA...640A..35R}, 104.\,\citet{2020ApJ...895...60R}, 105.\,\citet{2023NewAR..9601672K}, 106.\,\citet{2014Natur.505..378C}, 107.\,\citet{2020AA...640A..96P}, 108.\,\citet{2021Sci...371.1046M}, 109.\,\citet{2021ApJ...913...48G}, 110.\,\citet{2005MNRAS.364..899C}, 111.\,\citet{2001ApJ...555..489O},  112.\,\citet{2022ApJ...926..123A}.
}

\end{deluxetable*}
\end{longrotatetable}

\begin{figure}[htb!]
	\centering
	\includegraphics[scale=0.26]{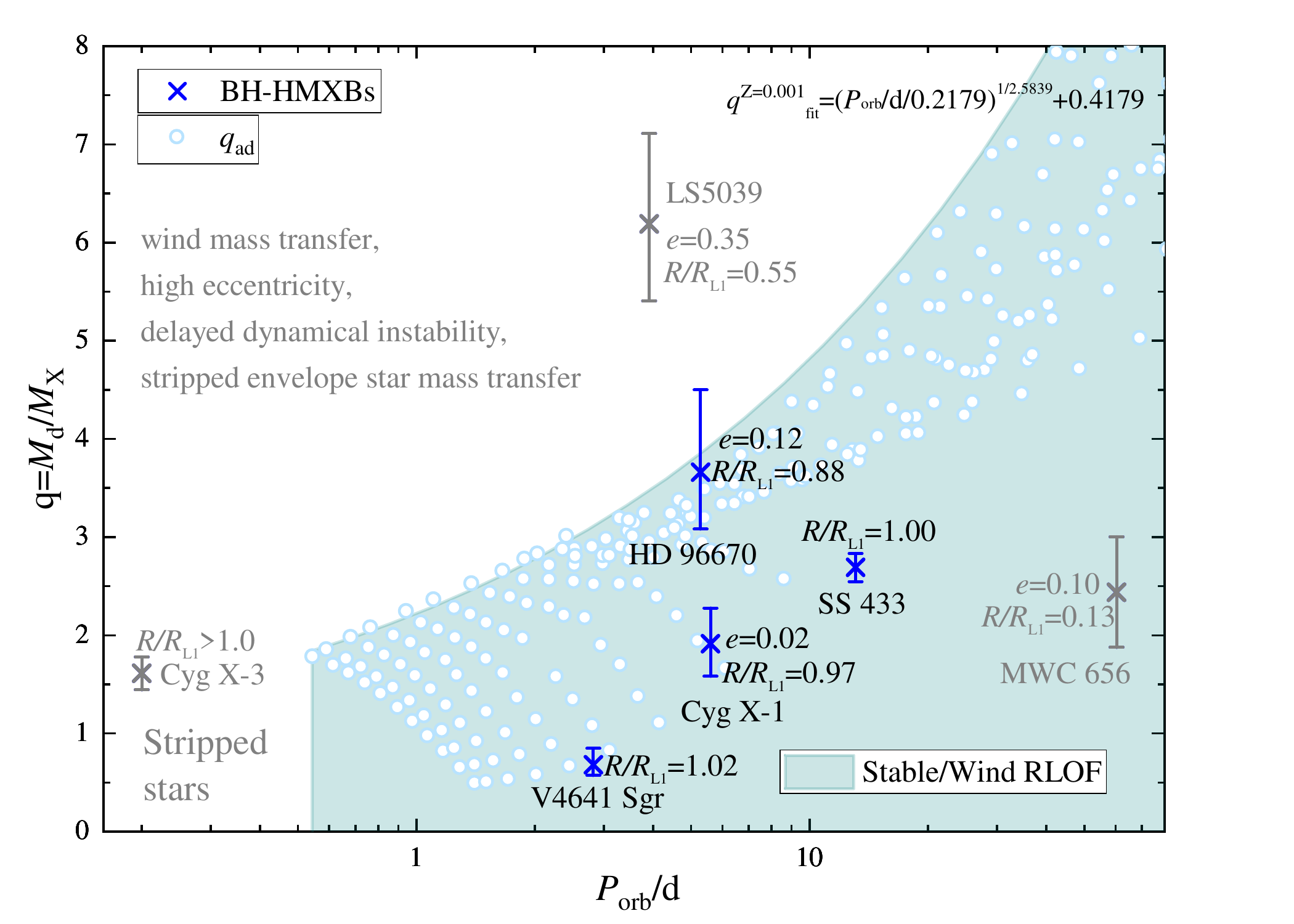}
	\caption{Seven BH-HMXBs in a mass ratio against the orbital period diagram. Blue symbols mark objects with a high overfilling factor $R/R_\mathrm{L}$ and low eccentricity $e$. Gray symbols show wind mass transferring objects or HMXBs with a near naked helium star donor. Blue open circles are the critical mass ratio for the dynamical timescale mass transfer of massive stars in different evolutionary states. The upper boundary of the blue region is a fit of the critical mass ratio as a function of the orbital period.}
	\label{fig-BH}
\end{figure}

Only 7 HMXBs composed with a BH accretor have known binary parameters  (Table\,\ref{tab-hmxbs}). We plot these BH-HMXBs in a mass ratio--orbital period diagram (Figure\,\ref{fig-BH}). Four have a low eccentricity and high overfilling factor (blue) and are suitable sources to compare with the theoretical mass ratio limits. MWC~656 and LS~5039 (gray) are wind-driven HMXBs. Cyg~X-3 contains a WN~4 to 6 type donor star which is significantly smaller than a massive main-sequence star. The blue open circles mark the critical mass ratio for massive metal-poor ($Z=0.001$) donor stars in different evolutionary states. We fit the maximum critical mass ratio $q_\mathrm{ad}$ envelope as a function of the orbital period. We only show the fitting of $Z=0.001$ massive stars for simplicity. The maximum critical mass ratios $q^\mathrm{max}_\mathrm{ad}$ of $Z=0.02$ massive stars are slightly larger than $Z=0.001$ donors. These observed BH-HMXBs in circular orbits and transferring mass through RLOF are located below our theoretically predicated mass ratio limits as required. 

Any systems above the blue envelope must be in a wind mass transfer phase or possibly undergoing thermal RLOF before experiencing a delayed dynamical instability. Indeed LS~5039 has a high eccentricity that is less correlation with RLOF.

\begin{figure*}[htb!]
	\centering
	\includegraphics[scale=0.55]{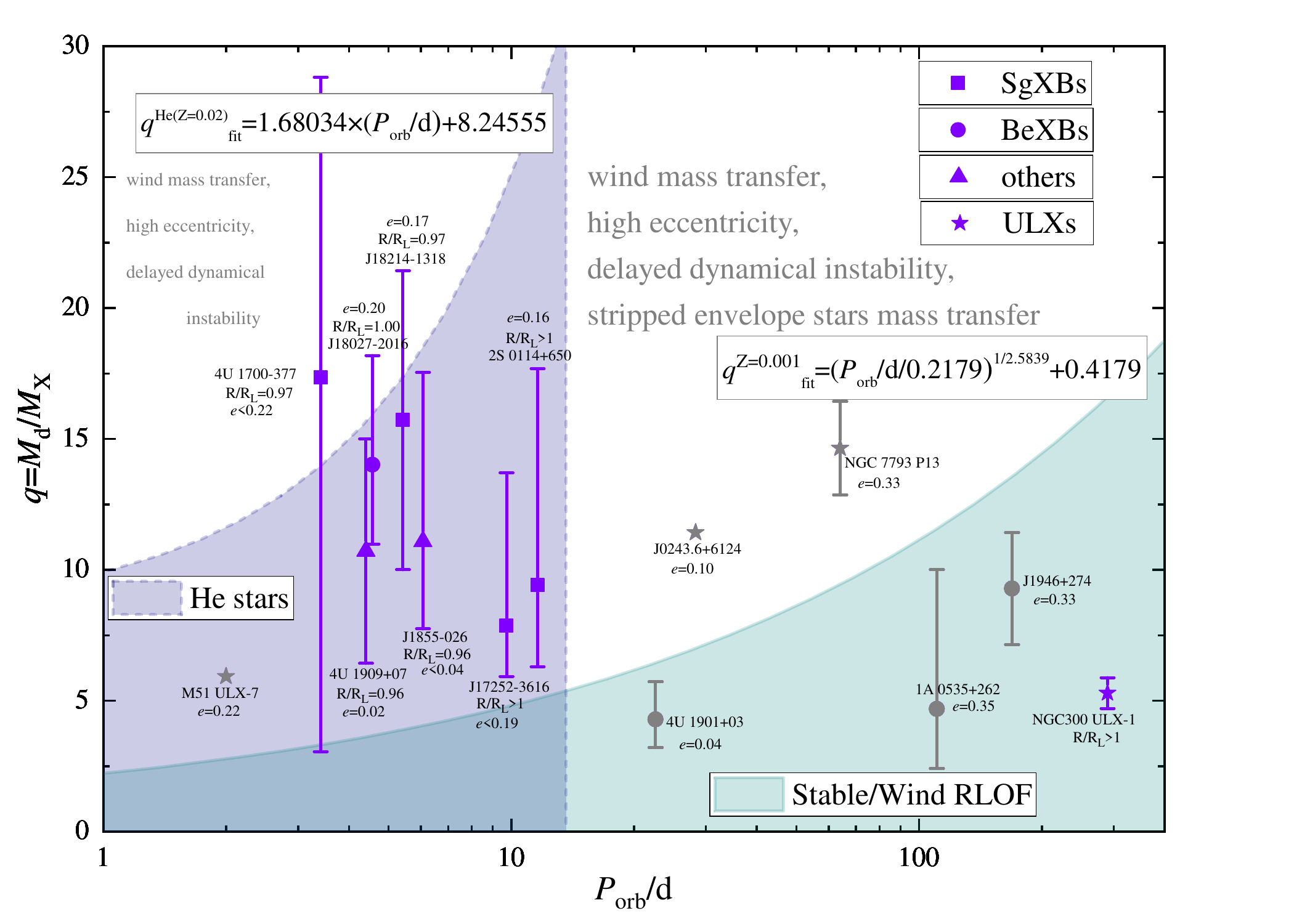}
	\caption{Fourteen NS-HMXBs on a mass ratio against orbital period diagram. Symbols mark different types of NS-HMXBs. Violet have measured radii while grey do not. The blue and violet regions are theoretically permitted spaces for stable RLOF from the normal and naked helium donors. Above the fitted maximum mass ratio envelope (blue for metal-poor $Z=0.001$ massive donors and violet for $Z=0.02$ helium donor stars), systems must be undergoing wind mass transfer, have a highly eccentric orbit, be suffering delayed dynamical instability. }
	\label{fig-NS}
\end{figure*}

Similarly we select 14 HMXB systems with an NS companions and eccentricity $e < 0.35$. We plot these NS-HMXBs in a mass ratio against orbital period diagram (Figure\,\ref{fig-NS}). Eight (violet) NS-HMXBs have known donor radii. The other six (grey) do not. We colour the stable mass transfer parameter space for low-metallicity ($Z=0.001$) hydrogen-rich donor stars blue in Figure\,\ref{fig-NS} similarly to Figure\,\ref{fig-BH}. So, above the blue envelope binary systems are only possible for wind mass transfer, thermal timescale mass transfer before entering delayed dynamically unstable mass transfer, or if the donor is a naked helium star. Our predicted mass ratio upper limits are for stable mass transfer in a circular orbit. With relatively long orbital periods ($P_\mathrm{orb} > 11\,{\rm d}$) NS-HMXBs, two eccentric systems are located above the critical mass ratio limit. These are most likely in a wind mass transfer phase. With relatively short orbital periods ($P_\mathrm{orb} < 11\,{\rm d}$), all eight systems located above the critical mass ratio limit. Seven have $R/R_\mathrm{L}$ larger than 0.96 and so must be in a state of relatively stable RLOF. They may be almost to experience a delayed dynamical instability. However, this stage is expected to last only for a short thermal timescale. A more likely explanation is that their donors are helium stars. We extrapolate the results of \citet{2024arXiv240613146Z} and fit the maximum critical mass ratios of solar-metallicity helium donor stars. We find these short-period NS-HMXBs are then under the critical mass ratio limit, except for one ready constrained eccentric system. We encourage more detailed observations to test this hypothesis.

\section{Discussions}	
\label{sec-discussion}

We have considered the impact of both metallicity and nonconservative mass transfer on the dynamical timescale mass transfer thresholds. Similarly to our previous study \citep{2023ApJ...945....7G}, metal-poor stars on the MS or HG (radiative envelope) tend to be more unstable than metal-rich donors (Figure\,\ref{fig-20m-qad}). Metal-poor donors, with deep convective envelopes, are more stable than metal-rich stars (Figure\,\ref{fig-20m-qad}). We expect that mass transfer stability thresholds may impact binary systems differently in the Galatic thick and thin disks and the halo. 

Nonconservative mass transfer significantly decreases the critical mass ratio for massive donors that have left the MS compared with the conservative case (Figure\,\ref{fig-massive-qad}). Nevertheless, the critical mass ratios of massive MS and HG donors are mostly still greater than 3 or 4. This suggests a moderate increase in the number of binary systems that form from stable mass transfer channels.

Stellar winds may significantly change the masses of the donor and the orbital separation before Roche-lobe filling. Such winds are included in binary population synthesis studies. Our initial donor stars that fill their Roche lobes are built without wind mass loss \citep{2010ApJ...717..724G} in order to minimize the impact of the uncertainty of stellar winds. Both observation \citep{2007A&A...465.1003M} and theory \citep{2008flhs.book.....C} suggest that stellar wind is metallicity-sensitive. We previously considered the impact of stellar wind \citep{2020ApJ...899..132G} and found that the critical mass ratio of a donor depends on the initial mass of the donor and its current core mass, both of which are evaluated in population synthesis models \citep{2002MNRAS.329..897H} in a different way.

Rotation is also important in massive star evolution. Rotation and extra mixing \citep{2003A&A...404..975M,2005A&A...429..581M} can enhance formation of Wolf-Rayet stars. Indeed, \citet{2004Sci...303..499S} noticed and \citet{2017ApJ...837..122M} reported a lackation of red supergiants in local group galaxies. So rotation and extra-mixing worthy of future work in mass transfer physics.

Our systematic studies of the thresholds for dynamical-timescale mass transfer predict the onset of the CE evolution \citep{1976IAUS...73...75P}. After the critical mass ratios for unstable mass transfer, the treatments of CE ejection is the next least understood question in binary star evolutio. The current treatment requires both precise binding energy $E_\mathrm{bind}$ \citep{2024ApJ...963L..35C} and CE ejection efficiency $\alpha_\mathrm{CE}$ \citep{2023MNRAS.518.3966S,2024ApJ...961..202G}. \citet{2022ApJ...933..137G,2024ApJ...961..202G} found evidence for a relation between the initial mass ratio and their common envelope ejection efficiency $\beta_\mathrm{CE}$ for low-mass ($M_\mathrm{i} < 2\,M_\odot$) donor stars. So this work should also be extended to higher mass stars.

%We also look forward to future studies on all kinds of binary systems using state-of-the-art binary population synthesis codes and physical inputs from binary evolution studies, such as COMPAS \citep{2022ApJS..258...34R,2023ApJ...958..138W,2024arXiv240611885R}, Binary\_C \citep{2023MNRAS.526.4130H}, METTISSE \citep{2023MNRAS.525..933A}, SEVN \citep{2023MNRAS.524..426I}, SeBa \citep{2024MNRAS.530.3706D}.

%Our theoretical studies will benifit from fast growing observation data for different binary star systems, such as WD binaries \citep{2021MNRAS.501.1677H,2022MNRAS.512.2625L,2023MNRAS.518.4579P,2024MNRAS.52711719Y,2024arXiv240506020Y}, double WDs \citep{2023MNRAS.526.5471Y,2024A&A...686A.221R}, hot subdwarf binaries \citep{2022A&A...666A.182S,2023A&A...673A..90S,2023NatAs...7..223L,2024NatAs...8..491L,2023MNRAS.524.3769W,2024A&A...684A.118U}, luminous red novae \citep{2021A&A...653A.134B}, Wolf-Rayle binaries \citep{2022ApJ...932...84M}, novae binaries \citep{2021MNRAS.504.6117K,2023MNRAS.523..305T}, binary pulsars \citep{2022MNRAS.514.4385S}, blue stragglers \citep{2021ApJ...908..229L,2024arXiv240520242N}, semidetached binaries \citep{2024ApJS..270...20X}, and stripped stars \citep{2021MNRAS.502.3436E}.

\section{Summary}
\label{sec-summary}

 Previously \citet{2010ApJ...717..724G,2015ApJ...812...40G,2020ApJ...899..132G} systematically presented critical mass ratios, assumed conserved mass and angular momentum, for dynamical-timescale mass transfer over the entire span of donor star evolutionary states with $Z=0.02$. Here we extend our analysis to include various metallicities and nonconservative mass transfer. Using $20\,M_\sun$ donor stars as examples, we analyze the response of stars with metallicities $Z=0.02$ and $Z=0.001$ with nonconservative mass transfer. We find the critical mass ratios of donor stars with masses between $1$ and $100\,M_\sun$ with $Z=0.001$. We further complement the thresholds for nonconservative unstable mass transfer for stars between $0.1$ and $100\,M_\sun$  at $Z=0.02$. Critical mass ratios of metal-poor MS and HG donors with radiative envelopes are smaller than those of solar metallicity stars at the same evolutionary stage. However, critical mass ratios of metal-poor RGB/AGB donors with convective envelope are larger than those of the solar metallicity with the same radii. Nonconservative mass transfer also decreases the critical mass ratios of donors from late MS to the end of the star's evolution. We apply our results to 82 observed HMXBs with measured mass ratios and orbital periods. The observed HMXBs with low eccentricities and overfilled Roche lobes are nicely located under our predicted limits. Our results can be applied to study individual binary objects and binary population synthesis of most binary systems.
 
\section{acknowledgments}
This project is supported by the National Natural Science Foundation of China (NSFC Nos. 12288102, 12125303, 12090040/3, 12173081), the National Key R\&D Program of China (Nos. 2021YFA1600403, 2021YFA1600401), the Key Research Program of Frontier Sciences, CAS (No. ZDBS-LY-7005), CAS-Light of West China Program, Yunnan Fundamental Research Projects (Nos. 202101AV070001, 202201BC070003, 202401BC070007), Yunnan Revitalization Talent Support Program - Science \& Technology Champion Project (No. 202305AB350003), and the International Centre of Supernovae, Yunnan Key Laboratory (No. 202302AN360001). HG thanks Professor Ronald Webbink for helpful discussions in building the adiabatic mass-loss model. HG thanks Boyuan Liu, a postdoctral research associate at institute of astronomy at university of cambridge, for his useful comments. CAT thanks Churchill College for his fellowship.

%\vspace{5mm}
%\facilities{}
%\software{}
\setcounter{table}{0}
\renewcommand{\thetable}{A\arabic{table}}
\appendix

\section{Tables for the initial physical properties of donor stars and the thresholds for dynamical timescale mass transfer} 
\label{sec:alltab}
We present the interior physical properties of metal-poor stars in Table\,\ref{tab-interior}. Only model sequences for $20\,M_\odot$ donors are listed in this printed table. Further models are available in machine-readable form. The physical parameters are\\
(1) $n$---mass-loss sequence number counting from ZAMS,\\
(2) $M_\mathrm{i}$---initial mass of the mass-transferring star ($M_\odot$),\\
(3) $\log_{10} t$---age (yr),\\
(4) $M_\mathrm{He}$---helium core mass ($M_\odot$), where the hydrogen abundance X first reaches 0.15 going out from the center,\\
(5) $M_\mathrm{C}$---carbon core mass ($M_\odot$) where the helium abundance Y first reaches 0.25,\\
(6) $M_\mathrm{deg}$---mass of the degenerate core ($M_\odot$),\\
(7) $M_\mathrm{ce}$---mass of the convective envelope ($M_\odot$),\\
(8) $\psi_\mathrm{c}$---central electron chemical potential ($\mu_\mathrm{e}$, in units of k$T$),\\
(9) $\log_{10} \rho_\mathrm{c}$---central density ( ${\rm g\,cm^{-3}}$),\\
(10) $\log_{10} T_\mathrm{c}$---central temperature (K),\\
(11) X$_\mathrm{c}$---central hydrogen mass fraction,\\
(12) Y$_\mathrm{c}$---central helium mass fraction, and\\ 
(13) X$_\mathrm{s}$---surface hydrogen mass fraction.

\begin{deluxetable*}{lllllllllllll}
	%\tabletypesize{\scriptsize}
	\tablewidth{0pt}
	\tablecolumns{13}
	%\rotate
	\tablecaption{Interior physical properties of metal-poor donor stars
		\label{tab-interior}}
	\tablehead{
		\colhead{$n$} & \colhead{$M_\mathrm{i}$} & \colhead{$\log_{10} t$} &\colhead{$M_\mathrm{He}$}
		&\colhead{$M_\mathrm{C}$} &\colhead{$M^{*}_\mathrm{deg}$} &\colhead{$M_\mathrm{ce}$}
		& \colhead{$\psi_\mathrm{c}$} & \colhead{$\log_{10} \rho_\mathrm{c}$} 
		&\colhead{$\log_{10} T_\mathrm{c}$} & \colhead{${\rm X_c}$} &\colhead{${\rm Y_c}$} &\colhead{${\rm X_s}$}\\
		\colhead{ } & \colhead{($M_\odot$)} & \colhead{(yr)} & \colhead{($M_\odot$)}
		&\colhead{($M_\odot$)} &\colhead{($M_\odot$)} & \colhead{($M_\odot$)}
		& \colhead{k$T$} & \colhead{($\mathrm{g\,cm^{-3}}$)} 
		&\colhead{(K)} & \colhead{ }& \colhead{ } & \colhead{ } \\
	}
	
	\startdata
	1	&	20.00 	&	4.29823** 	&	0.0000 	&	0.0000 	&	0.0000 	&	0.0000 	&	-5.813 	&	0.847 	&	7.602 	&	0.755 	&	0.244 	&	0.756 	\\
	2	&	20.00 	&	6.11811 	&	0.0000 	&	0.0000 	&	0.0000 	&	0.0000 	&	-5.867 	&	0.844 	&	7.606 	&	0.702 	&	0.297 	&	0.756 	\\
	3	&	20.00 	&	6.53654 	&	0.0000 	&	0.0000 	&	0.0000 	&	0.0000 	&	-5.964 	&	0.845 	&	7.617 	&	0.597 	&	0.402 	&	0.756 	\\
	4	&	20.00 	&	6.69735 	&	0.0000 	&	0.0000 	&	0.0000 	&	0.0000 	&	-6.047 	&	0.851 	&	7.627 	&	0.502 	&	0.497 	&	0.756 	\\
	5	&	20.00 	&	6.80333 	&	0.0000 	&	0.0000 	&	0.0000 	&	0.0000 	&	-6.129 	&	0.863 	&	7.639 	&	0.400 	&	0.599 	&	0.756 	\\
	6	&	20.00 	&	6.86733 	&	0.0000 	&	0.0000 	&	0.0000 	&	0.0000 	&	-6.195 	&	0.882 	&	7.651 	&	0.308 	&	0.691 	&	0.756 	\\
	7	&	20.00 	&	6.91894 	&	0.0000 	&	0.0000 	&	0.0000 	&	0.0000 	&	-6.254 	&	0.918 	&	7.668 	&	0.204 	&	0.795 	&	0.756 	\\
	8	&	20.00 	&	6.95590 	&	6.4784 	&	0.0000 	&	0.0000 	&	0.0000 	&	-6.279 	&	0.984 	&	7.693 	&	0.100 	&	0.899 	&	0.756 	\\
	9	&	20.00 	&	6.97635 	&	6.4439 	&	0.0000 	&	0.0000 	&	0.0000 	&	-6.203 	&	1.116 	&	7.738 	&	0.023 	&	0.976 	&	0.756 	\\
	10	&	20.00 	&	6.98154 	&	6.4464 	&	0.0000 	&	0.0000 	&	0.0000 	&	-5.404 	&	1.758 	&	7.926 	&	0.000 	&	0.999 	&	0.756 	\\
	11	&	20.00 	&	6.98180 	&	6.0535 	&	0.0000 	&	0.0000 	&	0.0000 	&	-4.663 	&	2.231 	&	8.025 	&	0.000 	&	0.999 	&	0.756 	\\
	12	&	20.00 	&	6.98194 	&	5.9880 	&	0.0000 	&	0.0000 	&	0.0000 	&	-4.293 	&	2.493 	&	8.091 	&	0.000 	&	0.999 	&	0.756 	\\
	13	&	20.00 	&	6.98207 	&	5.9508 	&	0.0000 	&	0.0000 	&	0.0000 	&	-3.998 	&	2.716 	&	8.153 	&	0.000 	&	0.999 	&	0.756 	\\
	14	&	20.00 	&	6.98219 	&	5.9264 	&	0.0000 	&	0.0000 	&	0.0000 	&	-3.810 	&	2.889 	&	8.213 	&	0.000 	&	0.998 	&	0.756 	\\
	15	&	20.00 	&	6.98230 	&	5.9143 	&	0.0000 	&	0.0000 	&	0.0000 	&	-3.803 	&	2.933 	&	8.239 	&	0.000 	&	0.995 	&	0.756 	\\
	16	&	20.00 	&	6.98243 	&	5.9102 	&	0.0000 	&	0.0000 	&	0.0000 	&	-3.841 	&	2.922 	&	8.243 	&	0.000 	&	0.991 	&	0.756 	\\
	17	&	20.00 	&	6.98260 	&	5.9129 	&	0.0000 	&	0.0000 	&	0.0000 	&	-3.868 	&	2.911 	&	8.243 	&	0.000 	&	0.986 	&	0.756 	\\
	18	&	20.00 	&	6.98305 	&	5.9355 	&	0.0000 	&	0.0000 	&	0.0000 	&	-3.887 	&	2.905 	&	8.244 	&	0.000 	&	0.972 	&	0.756 	\\
	19	&	20.00 	&	6.98667 	&	6.2021 	&	0.0000 	&	0.0000 	&	0.0000 	&	-3.939 	&	2.896 	&	8.253 	&	0.000 	&	0.859 	&	0.756 	\\
	20	&	20.00 	&	6.99093 	&	6.5194 	&	0.0000 	&	0.0000 	&	0.0000 	&	-3.986 	&	2.893 	&	8.264 	&	0.000 	&	0.730 	&	0.756 	\\
	21	&	20.00 	&	6.99534 	&	6.8261 	&	0.0000 	&	0.0000 	&	0.0000 	&	-4.024 	&	2.896 	&	8.277 	&	0.000 	&	0.598 	&	0.756 	\\
	22	&	20.00 	&	6.99987 	&	7.1116 	&	0.0000 	&	0.0000 	&	0.0000 	&	-4.048 	&	2.908 	&	8.291 	&	0.000 	&	0.467 	&	0.756 	\\
	23	&	20.00 	&	7.00351 	&	7.3153 	&	0.0000 	&	0.0000 	&	0.0000 	&	-4.055 	&	2.924 	&	8.304 	&	0.000 	&	0.366 	&	0.756 	\\
	24	&	20.00 	&	7.00612 	&	7.4467 	&	0.0000 	&	0.0000 	&	0.0000 	&	-4.053 	&	2.942 	&	8.314 	&	0.000 	&	0.296 	&	0.756 	\\
	25	&	20.00 	&	7.00787 	&	7.5275 	&	0.0000 	&	0.0000 	&	0.0002 	&	-4.047 	&	2.957 	&	8.322 	&	0.000 	&	0.252 	&	0.756 	\\
	26	&	20.00 	&	7.00895 	&	7.5746 	&	5.1621 	&	0.0000 	&	0.0115 	&	-4.042 	&	2.967 	&	8.327 	&	0.000 	&	0.225 	&	0.756 	\\
	27	&	20.00 	&	7.00952 	&	7.5989 	&	5.2912 	&	0.0000 	&	0.0322 	&	-4.038 	&	2.973 	&	8.330 	&	0.000 	&	0.212 	&	0.756 	\\
	28	&	20.00 	&	7.00990 	&	7.6143 	&	5.3466 	&	0.0000 	&	0.2851 	&	-4.036 	&	2.977 	&	8.332 	&	0.000 	&	0.204 	&	0.756 	\\
	29	&	20.00 	&	7.01166 	&	7.6771 	&	5.5674 	&	0.0000 	&	5.4279 	&	-4.020 	&	2.999 	&	8.342 	&	0.000 	&	0.164 	&	0.756 	\\
	30	&	20.00 	&	7.02019 	&	7.9333 	&	5.8987 	&	0.0000 	&	9.8211 	&	-3.204 	&	3.736 	&	8.585 	&	0.000 	&	0.000 	&	0.741 	\\
	\enddata
	\tablecomments{This table is available entirely in machine-readable form.}
	\tablecomments{* The degenerate core mass only makes sense for low- and intermediate-mass stars but not for massive stars. }
	\tablecomments{** The age of the ZAMS star is not zero because it contracts slightly to reach a minimum radius after core hydrogen burning.}
\end{deluxetable*}

Secondly, we document the global physical properties of metal-poor stars in Table\,\ref{tab-global}. As for Table\,\ref{tab-interior}, only model sequences of $20\,M_\odot$ donors are listed in this printed table. Again the full data are available online in machine-readable form. The variables are listed as follows:\\
(1) $n$---mass-loss sequence number,\\
(2) $M_\mathrm{i}$---initial mass of the mass-transferring star ($M_\odot$),\\
(3) $\log_{10} R_\mathrm{i}$---initial radius of the donor star ($R_\odot$),\\
(4) $\log_{10} T_\mathrm{eff}$---effective temperature (K),\\
(5) $\log_{10} L$---stellar luminosity ($L_\odot$),\\
(6) $\log_{10} t_\mathrm{KH}$—thermal, Kelvin--Helmholtz, timescale (yr),\\
(7) $k^2=I/(M_\mathrm{i}R_\mathrm{i}^2)$—dimensionless moment of inertia,\\
(8) $L_\mathrm{H}$---hydrogen-burning luminosity ($L_\odot$),\\
(9) $L_\mathrm{He}$---helium-burning luminosity ($L_\odot$),\\
(10) $L_\mathrm{Z}$---heavy-element (carbon-, oxygen- etc.) burning
luminosity ($L_\odot$),\\
(11) $|L_\mathrm{neu}|$---neutrino luminosity ($L_\odot$), and \\
(12) $L_\mathrm{th}$---gravothermal luminosity ($L_\odot$).

\begin{deluxetable*}{lllllllllllll}
	\tabletypesize{\scriptsize}
	\tablewidth{0pt}
	\tablecolumns{13}
	%\rotate
	\tablecaption{Global physical properties of metal-poor donor stars
		\label{tab-global}}
	\tablehead{
		\colhead{$n$} & \colhead{$M_\mathrm{i}$} & \colhead{$\log_{10} R_\mathrm{i}$} &\colhead{$\log_{10} T_\mathrm{eff}$}
		&\colhead{$\log_{10} L$} &\colhead{$\log_{10} t_\mathrm{KH}$} &\colhead{$k^2=I/(M_\mathrm{i}R_\mathrm{i}^2)$}
		& \colhead{$L_\mathrm{H}$} & \colhead{$L_\mathrm{He}$} 
		&\colhead{$L_\mathrm{Z}$} & \colhead{$|L_\mathrm{neu}|$} &\colhead{$L_\mathrm{th}$} \\
		\colhead{ } & \colhead{($M_\odot$)} & \colhead{($R_\odot$)} & \colhead{(K)}
		&\colhead{($L_\odot$)} &\colhead{(yr)} & \colhead{ }
		& \colhead{($L_\odot$)} & \colhead{($L_\odot$)} 
		&\colhead{($L_\odot$)} & \colhead{($L_\odot$)}& \colhead{($L_\odot$)}  \\
	}
	
	\startdata
	1 	&	20.00 	&	0.6619	&	4.5864	&	4.6227	&	4.8126	&	9.5611E-02	&	4.4605E+04	&	7.2782E-19	&	0.0000E+00	&	2.8758E+03	&	2.1898E+02	\\
	2 	&	20.00 	&	0.6829	&	4.5853	&	4.6604	&	4.7538	&	9.0493E-02	&	4.8692E+04	&	1.2382E-18	&	0.0000E+00	&	3.1314E+03	&	1.9047E+02	\\
	3 	&	20.00 	&	0.7264	&	4.5814	&	4.7317	&	4.6391	&	8.0609E-02	&	5.7370E+04	&	3.5298E-18	&	0.0000E+00	&	3.6785E+03	&	2.1697E+02	\\
	4 	&	20.00 	&	0.7703	&	4.5745	&	4.7921	&	4.5347	&	7.2018E-02	&	6.5934E+04	&	8.8420E-18	&	0.0000E+00	&	4.2192E+03	&	2.3954E+02	\\
	5 	&	20.00 	&	0.8246	&	4.5629	&	4.8540	&	4.4185	&	6.3049E-02	&	7.6044E+04	&	2.4612E-17	&	0.0000E+00	&	4.8586E+03	&	2.6624E+02	\\
	6 	&	20.00 	&	0.8804	&	4.5480	&	4.9062	&	4.3106	&	5.5405E-02	&	8.5746E+04	&	6.6279E-17	&	0.0000E+00	&	5.4733E+03	&	2.9407E+02	\\
	7 	&	20.00 	&	0.9533	&	4.5253	&	4.9608	&	4.1831	&	4.7336E-02	&	9.7253E+04	&	2.4640E-16	&	0.0000E+00	&	6.2035E+03	&	3.1832E+02	\\
	8 	&	20.00 	&	1.0364	&	4.4960	&	5.0102	&	4.0506	&	4.0028E-02	&	1.0897E+05	&	1.5946E-15	&	0.0000E+00	&	6.9483E+03	&	3.4505E+02	\\
	9 	&	20.00 	&	1.0886	&	4.4785	&	5.0443	&	3.9642	&	3.5284E-02	&	1.1759E+05	&	4.0070E-14	&	0.0000E+00	&	7.4971E+03	&	6.4378E+02	\\
	10 	&	20.00 	&	1.0230	&	4.5216	&	5.0858	&	3.9884	&	3.1714E-02	&	9.7016E+04	&	2.0580E-08	&	0.0000E+00	&	6.1882E+03	&	3.1005E+04	\\
	11 	&	20.00 	&	1.1214	&	4.4726	&	5.0864	&	3.8893	&	2.9389E-02	&	1.2727E+05	&	4.3474E-04	&	0.0000E+00	&	8.1177E+03	&	2.8640E+03	\\
	12 	&	20.00 	&	1.2227	&	4.4273	&	5.1078	&	3.7666	&	2.5969E-02	&	1.3177E+05	&	3.6568E-01	&	1.8090E-39	&	8.4045E+03	&	4.8165E+03	\\
	13 	&	20.00 	&	1.3212	&	4.3817	&	5.1225	&	3.6535	&	2.3269E-02	&	1.3212E+05	&	9.3762E+01	&	5.1829E-35	&	8.4266E+03	&	8.8000E+03	\\
	14 	&	20.00 	&	1.4203	&	4.3348	&	5.1332	&	3.5436	&	2.0915E-02	&	1.2911E+05	&	8.2745E+03	&	3.1811E-29	&	8.2368E+03	&	6.7365E+03	\\
	15 	&	20.00 	&	1.5191	&	4.2881	&	5.1437	&	3.4344	&	1.8612E-02	&	1.1810E+05	&	4.0547E+04	&	1.4221E-26	&	7.5366E+03	&	-1.1900E+04	\\
	16 	&	20.00 	&	1.6227	&	4.2400	&	5.1586	&	3.3158	&	1.6018E-02	&	1.1558E+05	&	4.7246E+04	&	9.6270E-26	&	7.3765E+03	&	-1.1380E+04	\\
	17 	&	20.00 	&	1.7289	&	4.1910	&	5.1749	&	3.1934	&	1.3439E-02	&	1.1701E+05	&	4.5724E+04	&	2.9139E-25	&	7.4677E+03	&	-5.6908E+03	\\
	18 	&	20.00 	&	1.8236	&	4.1471	&	5.1889	&	3.0846	&	1.1311E-02	&	1.1681E+05	&	4.5809E+04	&	1.4271E-24	&	7.4553E+03	&	-6.6743E+02	\\
	19 	&	20.00 	&	1.9235	&	4.1002	&	5.2009	&	2.9727	&	9.3513E-03	&	1.1231E+05	&	5.3882E+04	&	1.0912E-22	&	7.1697E+03	&	-1.9171E+02	\\
	20 	&	20.00 	&	2.0233	&	4.0529	&	5.2114	&	2.8625	&	7.6560E-03	&	1.0532E+05	&	6.4253E+04	&	1.3817E-21	&	6.7283E+03	&	-1.5158E+02	\\
	21 	&	20.00 	&	2.1219	&	4.0059	&	5.2207	&	2.7546	&	6.2239E-03	&	9.6559E+04	&	7.5929E+04	&	1.1673E-20	&	6.1754E+03	&	-1.0413E+02	\\
	22 	&	20.00 	&	2.2232	&	3.9573	&	5.2287	&	2.6453	&	5.0165E-03	&	8.6217E+04	&	8.8652E+04	&	9.1919E-20	&	5.5251E+03	&	-3.5717E+01	\\
	23 	&	20.00 	&	2.3241	&	3.9082	&	5.2341	&	2.5389	&	4.1074E-03	&	7.7279E+04	&	9.9131E+04	&	4.8220E-19	&	4.9660E+03	&	-2.8225E+00	\\
	24 	&	20.00 	&	2.4243	&	3.8589	&	5.2375	&	2.4354	&	3.5368E-03	&	7.0453E+04	&	1.0674E+05	&	1.6826E-18	&	4.5415E+03	&	1.1836E+02	\\
	25 	&	20.00 	&	2.5217	&	3.8107	&	5.2392	&	2.3363	&	3.2843E-03	&	6.5806E+04	&	1.1182E+05	&	4.0214E-18	&	4.2545E+03	&	9.2922E+01	\\
	26 	&	20.00 	&	2.6240	&	3.7595	&	5.2395	&	2.2336	&	3.3526E-03	&	6.2869E+04	&	1.1493E+05	&	6.9895E-18	&	4.0740E+03	&	-1.5093E+02	\\
	27 	&	20.00 	&	2.7242	&	3.7091	&	5.2381	&	2.1349	&	4.1434E-03	&	6.1136E+04	&	1.1658E+05	&	9.3677E-18	&	3.9674E+03	&	-7.3906E+02	\\
	28 	&	20.00 	&	2.8234	&	3.6566	&	5.2266	&	2.0471	&	1.2175E-02	&	5.8794E+04	&	1.1761E+05	&	1.1330E-17	&	3.8208E+03	&	-4.0798E+03	\\
	29 	&	20.00 	&	2.8730	&	3.6329	&	5.2309	&	1.9933	&	5.2887E-02	&	5.0393E+04	&	1.2256E+05	&	3.0375E-17	&	3.3009E+03	&	5.1427E+02	\\
	30 	&	20.00 	&	2.9294	&	3.6221	&	5.3003	&	1.8673	&	8.8615E-02	&	5.7779E+04	&	3.6313E+04	&	1.2399E-07	&	8.3165E+03	&	1.1390E+05	\\
	\enddata
	\tablecomments{Again this table is available entirely in machine-readable form.}
\end{deluxetable*}

Thirdly, we summarize the quantitative results of thresholds for both conservative and nonconservative dynamical-timescale mass transfer in Table\,\ref{tab-threshold}. The superscripts in Table\,\ref{tab-threshold} indicate $\beta \in{\{1,0.5,0\}}$. During the mass transfer process, a fraction ($1- \beta$) of mass is lost from the system and takes away the specific angular momentum of the accretor. Columns (4) to (10) list the dynamical mass transfer thresholds for standard stellar models, and columns (11) to (17) are for models with artificially isentropic convective envelopes (a $\sim$ on a variable top). Variables are \\
(1) $n$---mass-loss sequence number,\\
(2) $M_\mathrm{i}$---initial mass of the mass-transferring star ($M_\odot$),\\
(3) $\log_{10} R_\mathrm{i}$---initial radius of the donor star ($R_\odot$),\\
(4) $\zeta^{1}_\mathrm{ad}$---critical radius–mass exponent for conservative dynamical-timescale
mass transfer,\\
(5) $q^{1}_\mathrm{ad}$---critical initial mass ratio for conservative dynamical-timescale
mass transfer,\\
(6) $M^{1}_\mathrm{KH}$---mass threshold at which $\dot{M} = -M_\mathrm{i}/t_\mathrm{KH}$ and the tangent point for the minimum $q^{1}_\mathrm{ad}$,\\
(7) $q^{0.5}_\mathrm{ad}$---critical initial mass ratio for semiconservative dynamical-timescale mass transfer,\\
(8) $M^{0.5}_\mathrm{KH}$---mass threshold at which $\dot{M} = -M_\mathrm{i}/t_\mathrm{KH}$ and the tangent point for the minimum $q^{0.5}_\mathrm{ad}$,\\
(9) $q^{0}_\mathrm{ad}$---critical initial mass ratio for fully nonconservative dynamical-timescale
mass transfer,\\
(10) $M^{0}_\mathrm{KH}$---mass threshold at which $\dot{M} = -M_\mathrm{i}/t_\mathrm{KH}$ and the tangent point for the minimum $q^{0}_\mathrm{ad}$,\\
(11) $\tilde{\zeta}^{1}_\mathrm{ad}$---critical radius–mass exponent for conservative dynamical-timescale mass transfer,\\
(12) $\tilde{q}^{1}_\mathrm{ad}$---critical initial mass ratio for conservative dynamical-timescale mass transfer,\\
(13) $\tilde{M}^{1}_\mathrm{KH}$---mass threshold at which $\dot{M} = -M_\mathrm{i}/t_\mathrm{KH}$ and the tangent point for the minimum $q^{1}_\mathrm{ad}$,\\
(14) $\tilde{q}^{0.5}_\mathrm{ad}$---critical initial mass ratio for semiconservative dynamical-timescale mass transfer,\\
(15) $\tilde{M}^{0.5}_\mathrm{KH}$---mass threshold at which $\dot{M} = -M_\mathrm{i}/t_\mathrm{KH}$ and the tangent point for the minimum $q^{0.5}_\mathrm{ad}$,\\
(16) $\tilde{q}^{0}_\mathrm{ad}$---critical initial mass ratio for fully nonconservative dynamical-timescale mass transfer, and\\
(17) $\tilde{M}^{0}_\mathrm{KH}$---mass threshold at which $\dot{M} = -M_\mathrm{i}/t_\mathrm{KH}$ and the tangent point for the minimum $q^{0}_\mathrm{ad}$.

\begin{deluxetable*}{lllllllllllllllll}
	\tabletypesize{\scriptsize}
	\tablewidth{0pt}
	\tablecolumns{17}
	%\rotate
	\tablecaption{Thresholds for conservative to nonconservative dynamical-timescale mass transfer of $Z=0.001$ donor stars
		\label{tab-threshold}}
	\tablehead{
		\colhead{$n$} & \colhead{$M_\mathrm{i}$} & \colhead{$\log_{10} R_\mathrm{i}$} & \colhead{$\zeta^1_\mathrm{ad}$}
		& \colhead{$q^{1}_\mathrm{ad}$} & \colhead{$M^{1}_\mathrm{KH}$} & \colhead{$q^{0.5}_\mathrm{ad}$}
		& \colhead{$M^{0.5}_\mathrm{KH}$} & \colhead{$q^{0}_\mathrm{ad}$} 
		&\colhead{$M^{0}_\mathrm{KH}$} & \colhead{$\tilde{\zeta}^1_\mathrm{ad}$}
		& \colhead{$\tilde{q}^1_\mathrm{ad}$} & \colhead{$\tilde{M}^1_\mathrm{KH}$} & \colhead{$\tilde{q}^{0.5}_\mathrm{ad}$}
		& \colhead{$\tilde{M}^{0.5}_\mathrm{KH}$} & \colhead{$\tilde{q}^{0}_\mathrm{ad}$} 
		& \colhead{$\tilde{M}^{0}_\mathrm{KH}$}\\
		\colhead{ } & \colhead{($M_\odot$)} & \colhead{($R_\odot$)} & \colhead{ }
		& \colhead{ } &\colhead{($M_\odot$)} & \colhead{ }
		& \colhead{($M_\odot$)} & \colhead{ } 
		& \colhead{($M_\odot$)} & \colhead{ }
		& \colhead{ } &\colhead{($M_\odot$)} & \colhead{ }
		& \colhead{($M_\odot$)} & \colhead{ } 
		& \colhead{($M_\odot$)}\\
	}
	
	\startdata
	1	&	20.00	&	0.6619	&	1.576	&	1.520	&	12.572	&	1.638	&	12.082	&	1.687	&	10.903	&	1.576	&	1.519	&	12.579	&	1.638	&	12.067	&	1.687	&	10.923	\\
	2	&	20.00	&	0.6829	&	1.702	&	1.578	&	12.723	&	1.693	&	12.224	&	1.738	&	11.115	&	1.702	&	1.578	&	12.723	&	1.693	&	12.224	&	1.738	&	11.115	\\
	3	&	20.00	&	0.7264	&	1.969	&	1.703	&	12.952	&	1.808	&	12.485	&	1.835	&	11.520	&	1.969	&	1.703	&	12.952	&	1.808	&	12.485	&	1.835	&	11.520	\\
	4	&	20.00	&	0.7703	&	2.235	&	1.826	&	13.128	&	1.921	&	12.667	&	1.929	&	11.746	&	2.233	&	1.826	&	13.122	&	1.921	&	12.660	&	1.929	&	11.756	\\
	5	&	20.00	&	0.8246	&	2.557	&	1.976	&	13.330	&	2.057	&	12.851	&	2.042	&	11.925	&	2.557	&	1.977	&	13.345	&	2.056	&	12.840	&	2.041	&	11.864	\\
	6	&	20.00	&	0.8804	&	2.886	&	2.130	&	13.531	&	2.194	&	13.054	&	2.152	&	12.024	&	2.886	&	2.130	&	13.532	&	2.194	&	13.048	&	2.152	&	12.010	\\
	7	&	20.00	&	0.9533	&	3.308	&	2.327	&	13.716	&	2.368	&	13.239	&	2.291	&	12.172	&	3.308	&	2.327	&	13.716	&	2.368	&	13.239	&	2.291	&	12.172	\\
	8	&	20.00	&	1.0364	&	3.787	&	2.551	&	13.924	&	2.562	&	13.408	&	2.442	&	12.277	&	3.788	&	2.552	&	13.924	&	2.562	&	13.408	&	2.442	&	12.277	\\
	9	&	20.00	&	1.0886	&	4.172	&	2.732	&	14.051	&	2.716	&	13.562	&	2.560	&	12.392	&	4.172	&	2.731	&	14.072	&	2.715	&	13.557	&	2.560	&	12.396	\\
	10	&	20.00	&	1.0230	&	4.479	&	2.876	&	14.300	&	2.842	&	13.764	&	2.663	&	12.689	&	4.479	&	2.876	&	14.300	&	2.842	&	13.764	&	2.663	&	12.689	\\
	11	&	20.00	&	1.1214	&	4.552	&	2.910	&	14.343	&	2.873	&	13.870	&	2.692	&	12.870	&	4.553	&	2.910	&	14.342	&	2.874	&	13.869	&	2.693	&	12.869	\\
	12	&	20.00	&	1.2227	&	5.022	&	3.130	&	14.380	&	3.053	&	13.876	&	2.818	&	12.779	&	5.028	&	3.133	&	14.379	&	3.055	&	13.874	&	2.819	&	12.776	\\
	13	&	20.00	&	1.3212	&	5.481	&	3.346	&	14.389	&	3.224	&	13.855	&	2.933	&	12.684	&	5.492	&	3.351	&	14.385	&	3.228	&	13.851	&	2.935	&	12.678	\\
	14	&	20.00	&	1.4203	&	5.965	&	3.573	&	14.379	&	3.402	&	13.815	&	3.048	&	12.571	&	5.983	&	3.582	&	14.373	&	3.408	&	13.808	&	3.052	&	12.562	\\
	15	&	20.00	&	1.5191	&	6.536	&	3.842	&	14.407	&	3.609	&	13.804	&	3.183	&	12.498	&	6.562	&	3.854	&	14.400	&	3.618	&	13.795	&	3.188	&	12.485	\\
	16	&	20.00	&	1.6227	&	7.335	&	4.218	&	14.476	&	3.895	&	13.850	&	3.368	&	12.454	&	7.376	&	4.238	&	14.466	&	3.909	&	13.839	&	3.375	&	12.438	\\
	17	&	20.00	&	1.7289	&	8.344	&	4.694	&	14.548	&	4.248	&	13.887	&	3.587	&	12.406	&	8.408	&	4.725	&	14.536	&	4.269	&	13.872	&	3.598	&	12.383	\\
	18	&	20.00	&	1.8236	&	9.408	&	5.197	&	14.611	&	4.609	&	13.918	&	3.803	&	12.360	&	9.502	&	5.242	&	14.594	&	4.638	&	13.899	&	3.818	&	12.332	\\
	19	&	20.00	&	1.9235	&	10.622	&	5.772	&	14.670	&	5.009	&	13.938	&	4.034	&	12.351	&	10.757	&	5.836	&	14.651	&	5.050	&	13.916	&	4.054	&	12.319	\\
	20	&	20.00	&	2.0233	&	12.014	&	6.432	&	14.725	&	5.455	&	13.955	&	4.280	&	12.340	&	12.206	&	6.523	&	14.702	&	5.511	&	13.929	&	4.306	&	12.302	\\
	21	&	20.00	&	2.1219	&	13.616	&	7.193	&	14.790	&	5.950	&	13.967	&	4.541	&	12.311	&	13.885	&	7.321	&	14.759	&	6.026	&	13.937	&	4.574	&	12.258	\\
	22	&	20.00	&	2.2232	&	15.506	&	8.092	&	14.845	&	6.514	&	13.973	&	4.821	&	12.257	&	15.882	&	8.271	&	14.815	&	6.616	&	13.939	&	4.863	&	12.204	\\
	23	&	20.00	&	2.3241	&	17.579	&	9.080	&	14.873	&	7.108	&	13.969	&	5.097	&	12.184	&	18.108	&	9.333	&	14.840	&	7.246	&	13.924	&	5.150	&	12.122	\\
	24	&	20.00	&	2.4243	&	19.752	&	10.118	&	14.924	&	7.708	&	13.986	&	5.361	&	12.105	&	20.875	&	10.655	&	14.858	&	7.989	&	13.915	&	5.463	&	12.001	\\
	25	&	20.00	&	2.5217	&	22.012	&	11.199	&	14.948	&	8.311	&	13.987	&	5.613	&	12.020	&	24.449	&	12.366	&	14.848	&	8.895	&	13.864	&	5.812	&	11.805	\\
	26	&	20.00	&	2.6240	&	24.641	&	12.458	&	14.791	&	8.985	&	14.005	&	5.878	&	11.911	&	29.519	&	14.800	&	14.803	&	10.093	&	13.790	&	6.225	&	11.574	\\
	27	&	20.00	&	2.7242	&	27.470	&	13.816	&	14.924	&	9.678	&	14.024	&	6.139	&	11.749	&	37.011	&	18.407	&	14.746	&	11.719	&	13.718	&	6.719	&	11.236	\\
	28	&	20.00	&	2.8234	&	29.117	&	14.607	&	14.946	&	10.100	&	14.193	&	6.362	&	11.866	&	46.343	&	22.914	&	14.724	&	13.622	&	13.819	&	7.282	&	10.962	\\
	29	&	20.00	&	2.8730	&	13.015	&	6.907	&	16.106	&	6.048	&	15.763	&	5.044	&	15.176	&	23.739	&	12.026	&	15.673	&	9.102	&	15.246	&	6.484	&	14.278	\\
	30	&	20.00	&	2.9294	&	10.891	&	5.900	&	14.318	&	5.058	&	13.672	&	4.041	&	12.411	&	18.087	&	9.323	&	13.728	&	7.013	&	12.942	&	4.887	&	11.655	\\
	\enddata
	\tablecomments{Similarly this table is available entirely in machine-readable form.}
\end{deluxetable*}

Finally, we complement the thresholds for nonconservative unstable mass transfer in $Z=0.02$ stars in Table\,\ref{tab-threshold2}. Variables are the same as in Table\,\ref{tab-threshold}. The full model grids cover 1567 ($Z=0.02$) donors with masses from 0.1 to 100\,$M_\odot$. \citet{2020ApJ...899..132G} documented these donor stars' corresponding initial physical parameters in their tables 1 and 2.

\begin{deluxetable*}{lllllllllllllllll}
	\tabletypesize{\scriptsize}
	\tablewidth{0pt}
	\tablecolumns{17}
	%\rotate
	\tablecaption{Thresholds for conservative to nonconservative dynamical-timescale mass transfer of $Z=0.02$ donor stars
		\label{tab-threshold2}}
	\tablehead{
		\colhead{$n$} & \colhead{$M_\mathrm{i}$} & \colhead{$\log_{10} R_\mathrm{i}$} & \colhead{$\zeta^1_\mathrm{ad}$}
		& \colhead{$q^{1}_\mathrm{ad}$} & \colhead{$M^{1}_\mathrm{KH}$} & \colhead{$q^{0.5}_\mathrm{ad}$}
		& \colhead{$M^{0.5}_\mathrm{KH}$} & \colhead{$q^{0}_\mathrm{ad}$} 
		&\colhead{$M^{0}_\mathrm{KH}$} & \colhead{$\tilde{\zeta}^1_\mathrm{ad}$}
		& \colhead{$\tilde{q}^1_\mathrm{ad}$} & \colhead{$\tilde{M}^1_\mathrm{KH}$} & \colhead{$\tilde{q}^{0.5}_\mathrm{ad}$}
		& \colhead{$\tilde{M}^{0.5}_\mathrm{KH}$} & \colhead{$\tilde{q}^{0}_\mathrm{ad}$} 
		& \colhead{$\tilde{M}^{0}_\mathrm{KH}$}\\
		\colhead{ } & \colhead{($M_\odot$)} & \colhead{($R_\odot$)} & \colhead{ }
		& \colhead{ } &\colhead{($M_\odot$)} & \colhead{ }
		& \colhead{($M_\odot$)} & \colhead{ } 
		& \colhead{($M_\odot$)} & \colhead{ }
		& \colhead{ } &\colhead{($M_\odot$)} & \colhead{ }
		& \colhead{($M_\odot$)} & \colhead{ } 
		& \colhead{($M_\odot$)}\\
	}
	
	\startdata
	1	&	20.00	&	0.7790	&	1.896	&	1.669	&	12.015	&	1.765	&	11.483	&	1.777	&	10.291	&	1.904	&	1.672	&	12.004	&	1.769	&	11.472	&	1.779	&	10.278	\\
	2	&	20.00	&	0.8239	&	2.198	&	1.809	&	12.264	&	1.893	&	11.847	&	1.885	&	10.553	&	2.211	&	1.815	&	12.248	&	1.898	&	11.833	&	1.890	&	10.029	\\
	3	&	20.00	&	0.8777	&	2.563	&	1.980	&	12.510	&	2.044	&	12.021	&	2.007	&	10.717	&	2.586	&	1.990	&	12.491	&	2.053	&	12.004	&	2.013	&	10.665	\\
	4	&	20.00	&	0.9360	&	2.972	&	2.170	&	12.701	&	2.210	&	12.222	&	2.135	&	10.763	&	3.008	&	2.187	&	12.675	&	2.224	&	12.198	&	2.144	&	10.679	\\
	5	&	20.00	&	1.0066	&	3.484	&	2.410	&	12.907	&	2.416	&	12.403	&	2.287	&	10.751	&	3.543	&	2.437	&	12.874	&	2.438	&	12.366	&	2.300	&	10.646	\\
	6	&	20.00	&	1.0940	&	4.151	&	2.722	&	13.110	&	2.676	&	12.570	&	2.471	&	10.708	&	4.244	&	2.765	&	13.070	&	2.710	&	12.514	&	2.490	&	10.589	\\
	7	&	20.00	&	1.1909	&	4.943	&	3.093	&	13.266	&	2.976	&	12.705	&	2.676	&	10.664	&	5.084	&	3.159	&	13.220	&	3.025	&	12.633	&	2.702	&	10.520	\\
	8	&	20.00	&	1.2612	&	5.682	&	3.440	&	13.402	&	3.248	&	12.799	&	2.859	&	10.776	&	5.876	&	3.531	&	13.348	&	3.314	&	12.721	&	2.893	&	10.623	\\
	9	&	20.00	&	1.1983	&	5.946	&	3.565	&	13.635	&	3.353	&	13.088	&	2.958	&	11.503	&	6.189	&	3.679	&	13.577	&	3.436	&	13.004	&	3.003	&	11.340	\\
	10	&	20.00	&	1.2973	&	6.268	&	3.716	&	13.671	&	3.469	&	13.122	&	3.040	&	11.801	&	6.524	&	3.836	&	13.608	&	3.555	&	13.048	&	3.088	&	11.676	\\
	11	&	20.00	&	1.3991	&	7.106	&	4.110	&	13.641	&	3.758	&	13.047	&	3.212	&	11.620	&	7.399	&	4.249	&	13.580	&	3.854	&	12.968	&	3.266	&	11.377	\\
	12	&	20.00	&	1.5013	&	8.001	&	4.532	&	13.647	&	4.057	&	12.951	&	3.380	&	11.431	&	8.332	&	4.689	&	13.566	&	4.161	&	12.885	&	3.433	&	11.293	\\
	13	&	20.00	&	1.6029	&	8.939	&	4.976	&	13.690	&	4.365	&	12.911	&	3.546	&	11.247	&	9.306	&	5.149	&	13.617	&	4.476	&	12.824	&	3.599	&	11.089	\\
	14	&	20.00	&	1.7035	&	9.923	&	5.441	&	13.765	&	4.682	&	12.915	&	3.711	&	11.034	&	10.326	&	5.632	&	13.705	&	4.801	&	12.836	&	3.765	&	10.924	\\
	15	&	20.00	&	1.8074	&	10.996	&	5.949	&	13.864	&	5.023	&	12.956	&	3.884	&	10.845	&	11.445	&	6.162	&	13.809	&	5.152	&	12.877	&	3.939	&	10.730	\\
	16	&	20.00	&	1.9068	&	12.090	&	6.468	&	13.990	&	5.366	&	13.011	&	4.057	&	10.681	&	12.594	&	6.707	&	13.929	&	5.506	&	12.934	&	4.113	&	10.568	\\
	17	&	20.00	&	2.0087	&	13.301	&	7.043	&	14.114	&	5.738	&	13.099	&	4.240	&	10.556	&	13.918	&	7.337	&	14.048	&	5.906	&	13.005	&	4.304	&	10.429	\\
	18	&	20.00	&	2.1065	&	14.586	&	7.654	&	14.236	&	6.125	&	13.168	&	4.424	&	10.462	&	15.306	&	7.997	&	14.168	&	6.316	&	13.074	&	4.494	&	10.311	\\
	19	&	20.00	&	2.2163	&	16.197	&	8.421	&	14.377	&	6.597	&	13.249	&	4.641	&	10.350	&	17.052	&	8.829	&	14.306	&	6.817	&	13.161	&	4.717	&	10.181	\\
	20	&	20.00	&	2.3061	&	17.590	&	9.086	&	14.494	&	6.998	&	13.319	&	4.820	&	10.272	&	18.747	&	9.638	&	14.409	&	7.287	&	13.214	&	4.915	&	10.072	\\
	21	&	20.00	&	2.4227	&	19.332	&	9.917	&	14.668	&	7.499	&	13.454	&	5.047	&	10.258	&	21.398	&	10.905	&	14.537	&	7.999	&	13.280	&	5.200	&	9.924	\\
	22	&	20.00	&	2.5158	&	20.777	&	10.608	&	14.706	&	7.907	&	13.585	&	5.232	&	10.258	&	24.051	&	12.176	&	14.573	&	8.681	&	13.331	&	5.456	&	9.759	\\
	23	&	20.00	&	2.6188	&	22.409	&	11.389	&	14.473	&	8.334	&	13.745	&	5.435	&	10.298	&	27.870	&	14.008	&	14.495	&	9.598	&	13.424	&	5.779	&	9.513	\\
	24	&	20.00	&	2.7163	&	23.755	&	12.034	&	14.472	&	8.654	&	13.790	&	5.620	&	10.490	&	32.284	&	16.130	&	14.164	&	10.564	&	13.495	&	6.116	&	9.266	\\
	25	&	20.00	&	2.8192	&	13.181	&	6.986	&	17.440	&	6.436	&	17.241	&	5.357	&	12.260	&	22.642	&	11.501	&	17.092	&	9.459	&	16.842	&	6.084	&	11.897	\\
	26	&	20.00	&	2.9088	&	8.333	&	4.689	&	15.672	&	4.365	&	15.254	&	3.856	&	14.437	&	14.213	&	7.477	&	14.875	&	6.197	&	14.492	&	4.825	&	13.480	\\
	27	&	20.00	&	2.9939	&	10.861	&	5.885	&	14.095	&	5.020	&	13.418	&	3.981	&	12.039	&	17.227	&	8.913	&	13.401	&	6.744	&	12.733	&	4.717	&	11.273	\\
	28	&	20.00	&	3.0828	&	26.315	&	13.261	&	11.691	&	8.519	&	10.992	&	5.153	&	10.178	&	81.139	&	39.808	&	11.016	&	16.928	&	10.444	&	7.207	&	9.754	\\
	\enddata
	\tablecomments{This table is available entirely in machine-readable form for 1567 stellar models. Only model sequences of $20\,M_\odot$ stars are presented here.}
	\tablecomments{ The initial physical properties of these stars are listed by \citet{2020ApJ...899..132G}.}
\end{deluxetable*}

\bibliography{hongwei8}{}
\bibliographystyle{aasjournal}

\listofchanges

\end{document}